\documentclass[aps,pra,reprint,showpacs,floatfix]{revtex4-1}
\usepackage{graphicx}
\usepackage{bm}
\usepackage{xcolor}
\usepackage{relsize}
\usepackage{amsmath}
\usepackage{float}
\usepackage[normalem]{ulem}

\begin{document}

\preprint{APS/123-QED}

\title{$\mathcal{PC}$-symmetry-protected edge states in interacting driven-dissipative bosonic systems}

\author{E. Cancellieri}
\email{emiliano.cancellieri@gmail.com}
\author{H. Schomerus}
\affiliation{Department of Physics, Lancaster University, Lancaster LA1 4YB, UK}
\date{\today}

\begin{abstract}
A main objective of topological photonics is the design of disorder-resilient optical devices. Many prospective applications would benefit from nonlinear effects, which not only are naturally present in real systems but also are needed for switching in computational processes, while the underlying particle interactions are a key ingredient for the manifestation of genuine quantum effects. A particularly attractive switching mechanism of dynamical systems are infinite-period bifurcations into limit cycles, as these set on with a finite amplitude. Here we describe how to realize this switching mechanism by combining attractive and repulsive particle interactions in a driven-dissipative Su-Schrieffer-Heeger model, such as realized in excitonic lasers and condensates so that the system displays a nonhermitian combination of parity and charge-conjugation ($\mathcal{PC}$) symmetry. We show that this symmetry survives in the nonlinear case and induces  infinite-period and limit-cycle bifurcations (distinct from a Hopf bifurcation) where the system switches from a symmetry-breaking stationary state into a symmetry-protected power-oscillating state of finite amplitude. These protected dynamical solutions display a number of characteristic features, among which are their finite amplitude at onset, their arbitrary long oscillation period close to threshold, and the symmetry of their frequency spectrum which provides a tuneable frequency comb. Phases with different transition scenarios are separated by exceptional points in the  stability spectrum, involving nonhermitian degeneracies of symmetry-protected excitations.
\end{abstract}

\maketitle

\section{Introduction}
Several recent experimental and theoretical proposals focus on the implementation of topologically non-trivial photonic systems, promising to make optical devices more efficient and resilient against disorder \cite{2018arXiv180204173O,Lu2014}. Much of this work focusses on the celebrated Su-Schrieffer-Heeger (SSH) model, originating in the description of electronic transport in conjugated polymers \cite{PhysRevLett.42.1698}, which can be implemented, e.g., using  photonic superlattices \cite{Malkova:09}, atom-optical lattices \cite{Atala2013}, microwave resonator arrays \cite{Poli2015},
microlaser arrays \cite{Zhao2018,PhysRevLett.120.113901,2018arXiv180401587Y,2018arXiv180609826O}, waveguide arrays \cite{PhysRevLett.115.040402}, plasmonic wave guides \cite{PhysRevB.96.045417}, as well as using the p-band of Bragg cavity-polariton pillars where the $p_x$ and the $p_y$ modes provide the alternated coupling between pillars in a zig-zag configuration \cite{St-Jean2017}.

As in the original electronic setting \cite{PhysRevLett.42.1698}, where topological defects are bound to solitonic lattice deformations, many of these platforms display nonlinear features. These may, e.g., be manifest in the lasing behavior associated with a nonlinear saturable pump \cite{Zhao2018,Malzard2018,Malzard2018a,laser_textbook,Zhou:16}, arise from nonlinear hopping terms that induce transitions between different topological phases \cite{PhysRevB.93.155112,1367-2630-19-9-095002,doi:10.1063/1.4976013}, induce the formation of self-localized states \cite{PhysRevLett.111.243905},  or take the form of strong blue-shifts of the photonic resonance due to exciton-exciton scattering \cite{St-Jean2017,RevModPhys.85.299}. These interactions can lead to dynamical edge instabilities in Bose condensates \cite{PhysRevA.88.063631}, induce
edge bound states for two particles \cite{PhysRevA.94.062704}, enable the tunability of edge states \cite{1805.03760}, and determine the stability of soliton-like solutions \cite{Gulevich2017,PhysRevA.90.023813,PhysRevLett.117.143901,PhysRevLett.96.063901,Suntsov:07,LEDERER20081} in open-dissipative scenarios with either resonant or non-resonant pump configurations.

In this work, we utilize a driven-dissipative extension of the SSH chain to demonstrate the dynamical switching  between robust stationary and nonstationary operation modes. The key ingredient is the inclusion of repulsive and attractive interactions distributed along the system. On each lattice site we account for interactions that either are repulsive and lead to an energy blue-shift, or are attractive and lead to an energy red-shift, as well as a nonlinear saturable pump. The interactions can conspire to break or preserve a dynamical symmetry in the system, where the latter case corresponds to balanced repulsive and attractive  interactions that alternate along the chain. The system then combines a parity symmetry $\mathcal{P}$ with a pseudo-spinful charge-conjugation symmetry $\mathcal{C}$, hence displays a nonhermitian $\mathcal{PC}$ symmetry squaring to $(\mathcal{PC})^2=-1$, whose consequences differ from the previously and extensively studied case of nonhermitian $\mathcal{PT}$ symmetry \cite{Bender,El-Ganainy:07,moiseyev_2011}, and in particular also extends dynamically to the nonlinear setting.

We show that this dynamical $\mathcal{PC}$ symmetry modifies the stability of the edge states to the extent that they can become unstable and undergo unconventional transitions from steady states to power-oscillating solutions that set in with a finite oscillation amplitude. We identify two distinct switching mechanisms by which the edge states become unstable, where the steady-state solutions can either coexist with power-oscillating solutions, or display a transition into power-oscillating solutions that initially have an infinite period. These power-oscillating states can be protected by the dynamical symmetry even in the situation when the finite linear system strictly does not admit topologically protected stationary states. The origin of the associated dynamical phase transitions can be traced back to exceptional points in the stability excitation spectrum.

This paper is structured as follows. In Section \ref{sec:model} we introduce the model, discuss its properties in the linear and closed setting, and describe how we extend it to the open-dissipative and nonlinear case. In Section \ref{sec:results1} we study the two cases of vanishing and uniform interactions as reference points to contrast with the switching mechanisms present for the configuration with balanced alternating interactions  that supports the dynamical $\mathcal{PC}$ symmetry, which is described in Section \ref{sec:results2}. In Section \ref{sec:conclusions} we present our conclusions and describe possible applications and extensions of this work. The appendices review the definition of a Bogoliubov excitation spectrum and place the $\mathcal{PC}$ symmetry encountered here into a wider context.

\section{Model\label{sec:model} and dynamical $\mathcal{PC}$ symmetry}
\begin{figure}
\centering
\includegraphics[scale = 0.45]{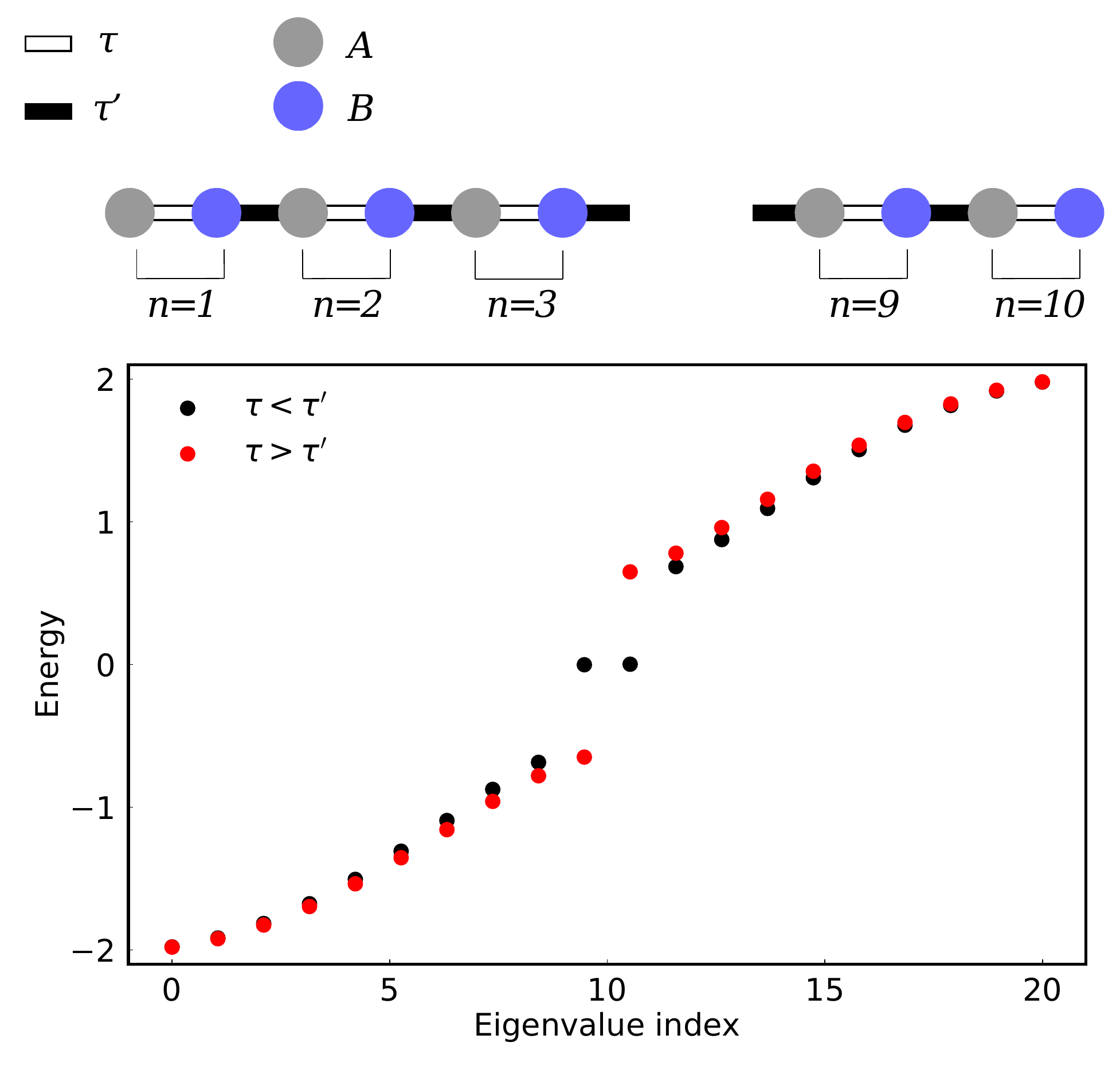}
\caption{{\bf Linear SSH model.} Top panel: sketch of a finite SSH chain with $N=10$ dimers (i.e. 20 sites). Bottom panel: energy levels for the linear closed system with couplings $\tau=0.7<\tau'=1.0$ (black) and $\tau=1.0>\tau'=0.7$ (red), ordered by magnitude. In the first case two weakly hybridized edge states appear at the center of the energy band gap. In this work we consider a driven-dissipative extension of this model which includes nonlinear interactions and saturable gain, as can be encountered in a polaritonic laser.}
\label{ssh}
\end{figure}

The SSH model is a tight-binding model for a one-dimensional dimer chain characterized by alternating strong and weak couplings between sites (see top of Fig.~\ref{ssh}). These alternating couplings define two distinct sublattices (A and B), representing the two sites on the dimer unit cell. The set of coupled mode equations are given by \begin{eqnarray}
i\frac{dA_n}{dt}&=&V_{n,A}(|A_n|^2)A_n+\tau B_n+\tau' B_{n-1},\nonumber
\\
i\frac{dB_n}{dt}&=&V_{n,B}(|B_n|^2)B_n+\tau A_n+\tau' A_{n+1},
\label{wave_equations}
\end{eqnarray}
where $A_n$ and $B_n$ are the amplitudes on the two sites on the $n$th dimer, $V_{n,s}$ ($s=A,B$) are effective onsite energies on each lattice site, and $\tau$ and $\tau'$ are the intra-dimer and inter-dimer couplings. The original linear setting of this model, $V_{n,s}=0$, is a periodic system with a symmetric band structure, where the two bands are separated by an energy gap of $\Delta=2|\tau-\tau'|$. One then can identify two topological phases depending whether $\tau>\tau'$ or $\tau<\tau'$, whose difference becomes apparent when one considers a semi-infinite chain. When the chain is terminated by a weak coupling ($\tau<\tau'$), one finds a symmetry-protected exponentially localized edge state with zero energy that only populates the A sublattice, and is absent when $\tau>\tau'$. In a finite system with an even number of sites and $\tau<\tau'$, a second edge states originates from the other edge, which is then localized on the B sublattice. These two edge states hybridize into a symmetric and an asymmetric solution with energy close to zero. This situation is illustrated in Fig.~\ref{ssh} for a chain with $N=10$ dimers.

Importantly, the energies of these hybridized states are not strictly protected by symmetry---they have a small, but nonvanishing energy, and do not constitute exact zero modes. As we will see, symmetry-protected periodically oscillating solution can however appear when nonlinear effects and in particular interactions are taken into account. In our nonlinear driven-dissipative extension of the SSH model, we therefore include a nonlinear complex effective potential
\begin{equation}
V_{n,s}(|s_n|^2)=g_{n,s}|s_n|^2+i\Gamma_{n,s}/(1+|s_n|^2)-i\gamma_{n,s},
\label{eq:v}
\end{equation}
where $g_{n,s}$ describes the on-site particle interactions (repulsive for $g_{n,s}>0$ and attractive for $g_{n,s}<0$), $\Gamma_{n,s}$ describes a nonlinear saturable pump, and $\gamma_{n,s}$  describes linear decay.

In absence of the interactions, this model describes the mode competition in a topological laser \cite{Malzard2018,Malzard2018a}, as realized in the recent experiments \onlinecite{Zhao2018,PhysRevLett.120.113901,2018arXiv180401587Y,2018arXiv180609826O}. The interactions themselves can break the symmetry protection of the system and shift the edge states away from their zero-mode position, as was exploited to characterize the exciton-polariton lasers of  Ref.~\onlinecite{St-Jean2017}. Here, we focus on the interplay of the nonlinear interaction and saturation effects, where for simplicity the decay rate $\gamma_{n,s}=\gamma$ is homogeneous, and contrast the cases where the interactions preserve, or break, the symmetry protection of the edge states of the linear system.

The case of symmetry-preserving interactions will be achieved by considering a balanced interaction scenario where  one sublattice displays repulsive interactions with $g_{n,A}=g>0$ while the other displays attractive interactions with $g_{n,B}=-g<0$. In this case, we find that the coupled mode equations \eqref{wave_equations} remain invariant under the substitution
\begin{equation}
A_n(t)\rightarrow B^*_{N-n+1}(t),\hspace{1cm}B_n(t)\rightarrow -A^*_{N-n+1}(t).
\label{eq:dynsym}
\end{equation}
This dynamical symmetry combines a nonhermitian charge-conjugation operation $\mathcal{C}$ \cite{PhysRevLett.115.200402} with a parity operation $\mathcal{P}$, so that the instantaneous Hamiltonian obeys $\mathcal{PC}H\mathcal{PC}=-H$ where $(\mathcal{PC})^2=-1$
(Appendix \ref{app:b} uses these features to place the $\mathcal{PC}$-symmetry into a wider context.)

For our version of the SSH model the $\mathcal{PC}$ symmetry  \eqref{eq:dynsym}
holds generally whenever the values of $\gamma_{n,s}$, $\Gamma_{n,s}$ are equal in symmetric positions of the chain with respect to its center, while  $g_{n,s}$ has opposite sign on these sites. We compare this balanced case with the case where all interactions are repulsive, corresponding to blue-shifts with $g_{n,s}=g>0$. For both scenarios we consider a pumping protocol on the terminating sites which preferably pumps the edge states ($\Gamma_{n,s}=\Gamma\ne 0$ only for $n=1,s=A$ and $n=N,s=B$).

When the dynamical $\mathcal{PC}$ symmetry \eqref{eq:dynsym} is respected, each stable solution with amplitudes $(A_n(t),B_n(t))$ on the $n$th dimer is paired with another solution where this dimer has amplitudes ($B^*_{N-n+1}(t),-A^*_{N-n+1}(t))$. Since this symmetry includes a complex-conjugation, the two paired solutions have opposite energy. Moreover, this symmetry allows the existence of self-symmetric solutions, which have a symmetric energy spectrum. The dynamical symmetry \eqref{eq:dynsym} therefore constitutes a natural reference point to separately explore the role of nonlinearities and symmetry-protection in an interacting driven-dissipative setting.

To study the physically stable solutions of the system in the presence of the nonlinear terms we numerically evaluate the time evolution of the coupled mode equations (\ref{wave_equations}) until a stationary state or an oscillating periodic solution are reached. Since several stable solutions may exist depending on the system parameters, this time evolution is performed for several different initial conditions. We characterize the solutions by two types of spectra: the frequency spectra $I_{A,B}(\omega)$ obtained by Fourier transformation of the time-dependent amplitudes on the two sublattices, and the complex Bogoliubov stability spectra $\omega_n$ obtained by linearization around the working point of the system (Appendix \ref{app:a} reviews the definition of this spectrum).

\section{Reference points\label{sec:results1}}
In this section we describe two reference points, with saturable gain but vanishing or uniformly repulsive interactions, to which we can then contrast our findings for the $\mathcal{PC}$-symmetric case of balanced interactions in the next Section.

\subsection{Noninteracting system with saturable gain}
\begin{figure}
\centering
\includegraphics[scale = 0.20]{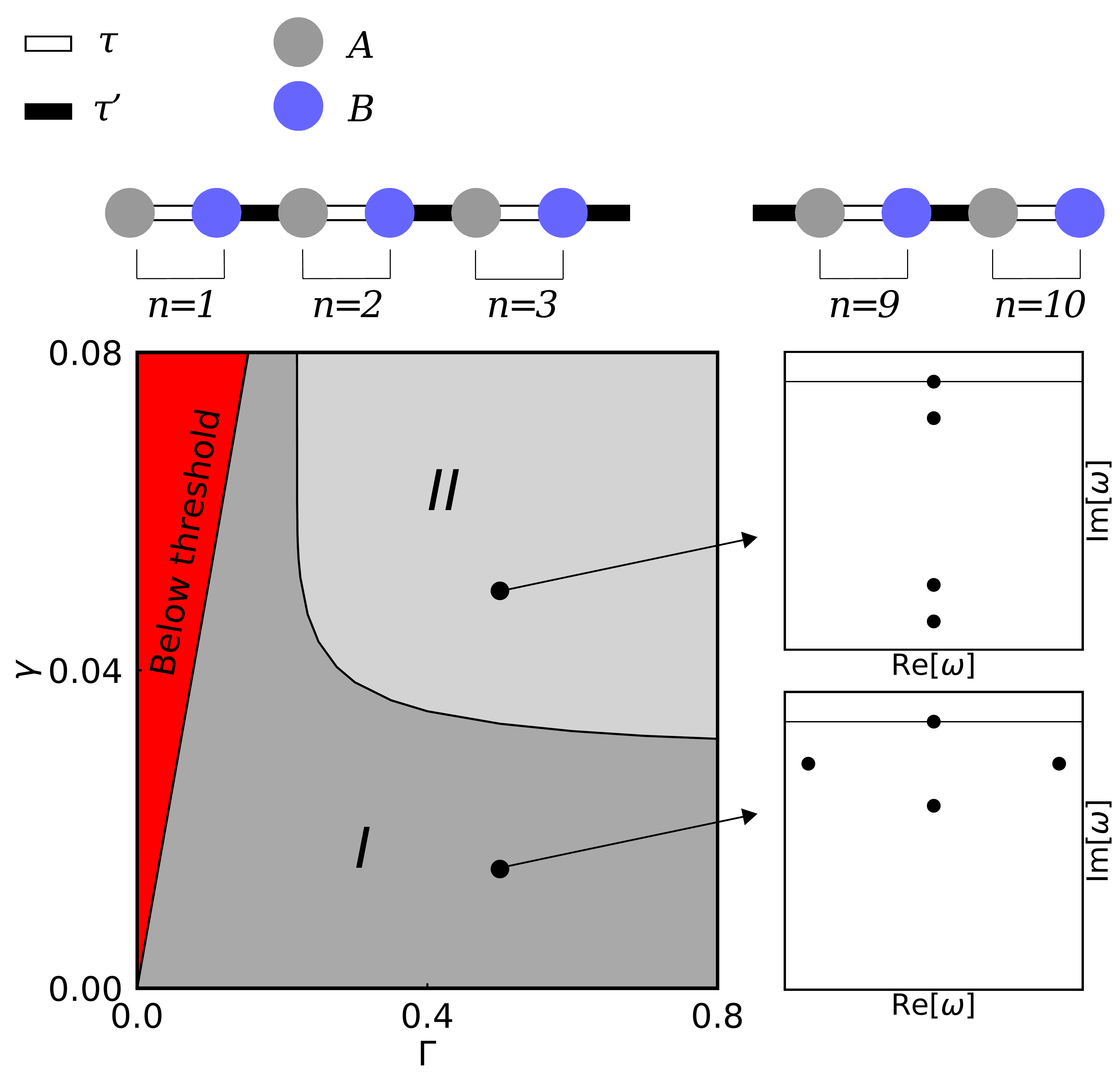}
\caption{{\bf Nonlinear pumped system with vanishing interactions \boldmath$g_{n,s}=0$.} Top: sketch of the chain with indication of the pumped sites. Left: phase diagram in the space of pump strength $\Gamma$ and decay rate $\gamma$, with couplings fixed to  $\tau'=0.7$ and $\tau=1.0$. In the red region the system is below threshold, so that its stationary state has vanishing intensity. In the gray  region the system supports two stationary states of finite intensity. The dark region I and the light region II differ by the configuration of the Bogoliubov excitation spectrum, which determines the stability of these stationary states. Two sample excitation spectra with  $\Gamma=0.5$ and $\gamma=0.015,0.050$ are shown in the right panels, where the horizontal black line indicates the real axis. At the transition between region I and II two excitations collide in an exceptional point on the imaginary axis.}
\label{zero_phase_space}
\end{figure}

As a first reference point we take a system with vanishing interactions $g_{n,s}=0$, where all nonlinearities occur due to the saturable pump. Figure \ref{zero_phase_space} shows the phase space for this case as a function of the intensity $\Gamma$ of the saturable pump and the linear decay rate $\gamma$. For small values of $\Gamma$ the intensity decays to zero as the loss rate is larger than the pump rate. Above this threshold, the system stabilizes in one of two possible stationary states with a finite intensity distribution, whose choice depends on the initial conditions. These states originate from the symmetric and antisymmetric hybridizations of the edge states in the linear regime, and are conjugate partners under the dynamical $\mathcal{PC}$ symmetry \eqref{eq:dynsym}. They therefore have conjugate frequency spectra, peaked at opposite positions close to zero frequency, and identical Bogoliubov spectra.

Further inspecting the Bogoliubov spectra as we vary $\gamma$ and $\Gamma$, we find that the phase space can be divided into two regions corresponding to two possible configurations, indicated as I and II. The first configuration occurs for small values of $\gamma$, where two of the four Bogoliubov eigenvalues closest to zero lie on the imaginary axis, while the other two lie symmetric with respect to it. For larger values of $\gamma$ this configuration changes in an exceptional point, after which all these four eigenvalues lie on the imaginary axis.

\subsection{Uniform interactions}
\begin{figure}
\centering
\includegraphics[scale = 0.28]
{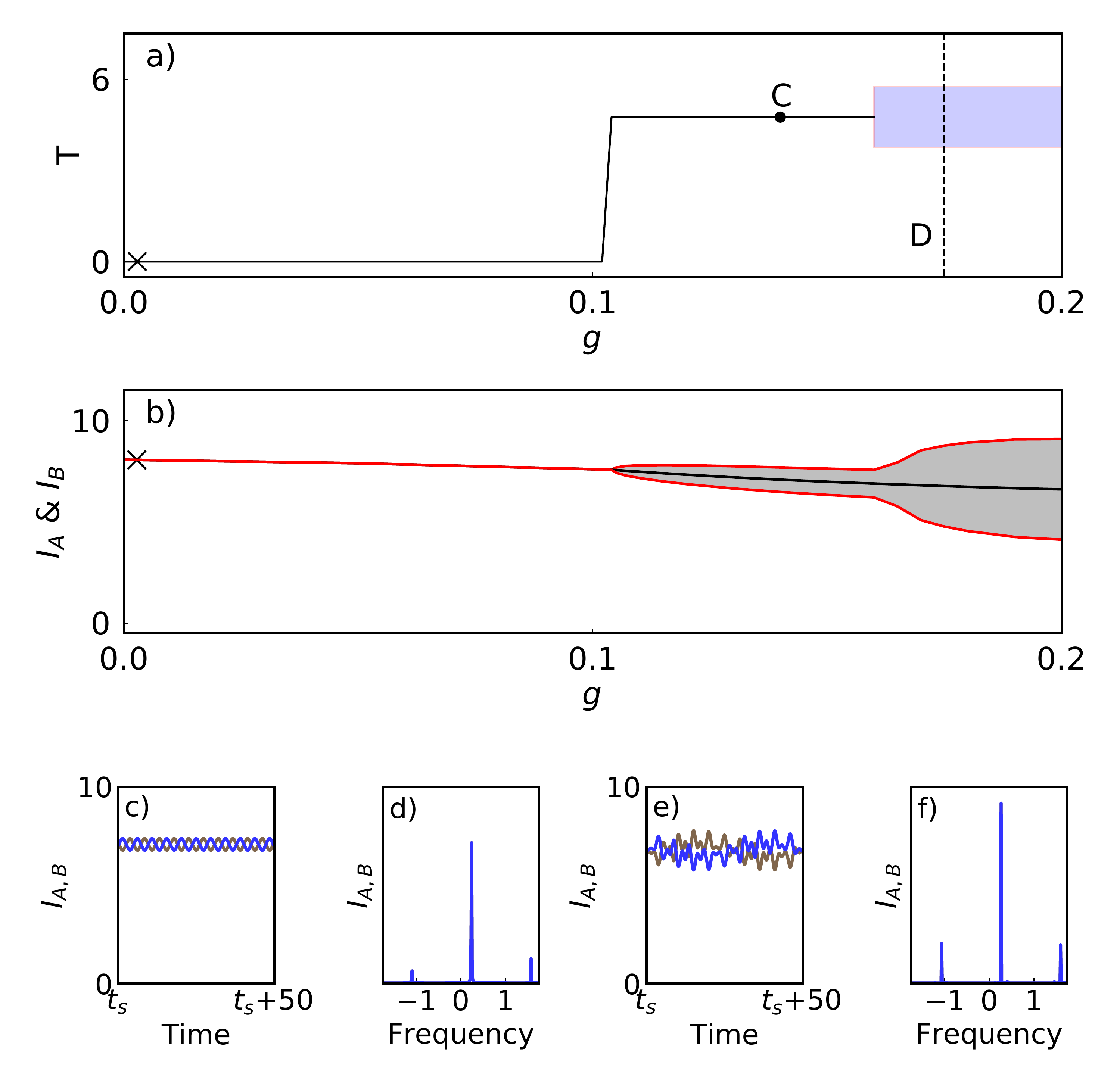}
\caption{{\bf Nonlinear pumped system with uniform interactions \boldmath$g_{n,s}=g>0$.} 
The interactions can destabilize the system, leading to power-oscillating states of period $T$.
Panel (a) shows the oscillation period $T$ and panel (b) the sublattice-resolved intensities $I_A,I_B$ as a function of $g$. The small cross at $g=0.0028$ indicates the threshold at which one of the spatially symmetric (bonding) stationary solutions destabilizes. The jump in period occurs when the second (antibonding) solution  destabilizes and is replaced by the power-oscillating state. The gray shaded region in panel (b) indicates the intensity oscillation amplitude, which is identical for $I_A$ and $I_B$ even though both intensities oscillate out of phase. 
Panels (c) and (d) show the time trace of the sublattice-resolved intensities $I_A$ (gray) and $I_B$ (blue) and the frequency spectrum at point C ($g=0.140$), while panels (e) and (f) show the analogous data for point D ($g=0.175$).
For all panels parameters are fixed to $\tau'=0.7$, $\tau=1.0$, $\gamma=0.05$, and $\Gamma_{1,A}=\Gamma_{10,B}=0.5$.}
\label{blue-shift}
\end{figure}

As a second reference point we consider a system with uniform interactions, which we assume to be repulsive so that $g_{n,s}=g>0$. These interactions break the chiral symmetry of the linear system in any nonuniform stationary state, and also break the dynamical $\mathcal{PC}$ symmetry given in Eq.~\eqref{eq:dynsym}. As shown in Fig.~\ref{blue-shift}, the interactions can destabilize the stationary states when we increase $g$ while fixing $\Gamma$ and $\gamma$. For small values of $g$ there are two stable solutions, which again originate from the bonding and antibonding states at $g=0$. Compared to this limit, however, the two frequency spectra are shifted up in energy, and no longer related to each other as the stationary and dynamical symmetries are broken. Further increasing the interactions, the solution with higher energy becomes unstable, where the threshold  depends on the pump and decay rates $\Gamma$ and $\gamma$. In the situation illustrated in the Fig.~\ref{blue-shift}, the threshold value $g=0.0028$ is very small, as indicated by the cross symbol in panels (a) and (b). The stationary state with the lower energy, however, remains stable for much larger interactions, up to $g_{th}=0.103$. Above this second threshold the system switches over into a power-oscillating mode, which sets on with a fixed period but initially small power-oscillation amplitude, which increases gradually from zero proportionally to $(g-g_{th})^{1/2}$. These features are indicative of a supercritical Hopf bifurcation \cite{Strogatz2015}. As shown in panels (c,d) for the point marked C, the corresponding frequency spectrum is slightly asymmetric but regularly structured, and includes a prominent central peak placed between two satellites that are situated close to the  energies of the former stationary solutions.

When we further increase the interaction strength we enter the gray region in Fig.~\ref{blue-shift}(a), where the oscillations become erratic, signifying the onset of chaos. This is illustrated in panels (e,f) for the solution marked D. We observe this scenario of consecutive instability across the whole parameter space (i.e., regardless whether we reside in region $I$ or $II$ in Figure \ref{zero_phase_space}).

These findings for the case of uniform interactions agree with the qualitative behaviour of a wide range of nonlinear optical systems.
In particular, the frequency spectra of the periodically oscillating solutions are very similar to the traditional optical parametric oscillator solutions in nonlinear crystals \cite{PhysRevLett.14.973} or in polariton cavities \cite{PhysRevB.71.115301,PhysRevB.92.035307,PhysRevLett.114.193901}, and their creation mechanism is generally associated to an Hopf bifurcation. As we now will show, much more versatile switching mechanisms can be realized for the $\mathcal{PC}$-symmetric case of balanced interactions.

\section{Balanced interactions\label{sec:results2}}
\begin{figure*}
\centering
\includegraphics[scale=0.33]{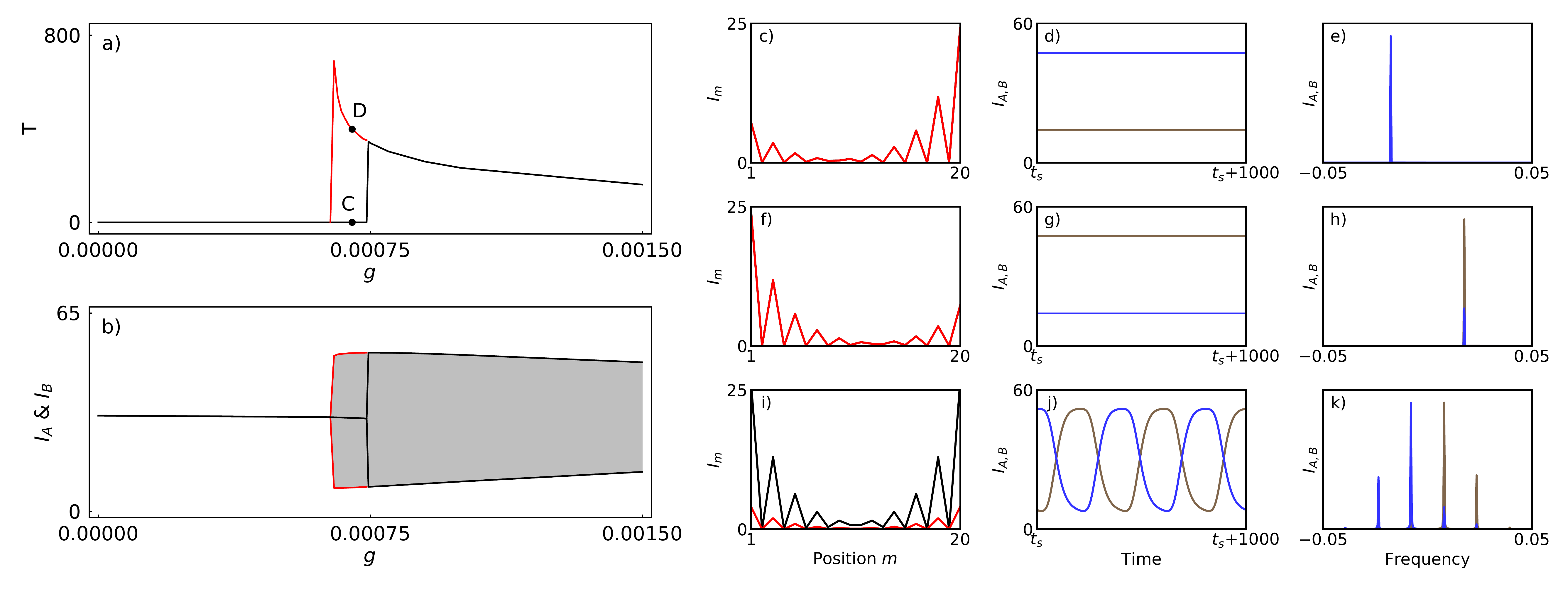}
\caption{{\bf Nonlinear pumped system with balanced interactions \boldmath$g_{n,A}=-g_{n,B}$
in region I.} Balanced interactions can drive the system into a power-oscillating state that is now protected by $\mathcal{PC}$-symmetry.
Panels (a) and (b) show the oscillation period $T$ and sublattice-resolved  intensities $I_A,I_B$ of stationary and oscillating states as a function of $g$.
The system displays a bistable interval with hysteretic behaviour between the stationary and power-oscillating solutions where the black branch with point C is followed when the interactions increase while the red branch with point D is followed when the interactions decrease, with both points positioned at $g=0.0007$. Panels (c-e) and (f-h) show the spatial intensity distribution, time trace of the sublattice-resolved intensities ($I_A$ in gray, $I_B$ in blue), and frequency spectrum of the two stationary states at point C. Panels (i-k) show the corresponding quantities at point D, where the black and red curves in the  intensity profile indicate the maximal and minimal intensity over an oscillation cycle. For all panels $\tau'=0.7$, $\tau=1.0$, $\gamma=0.015$, and $\Gamma=0.5$, corresponding to the point marked in region I of Fig.~\ref{zero_phase_space}.}
\label{oscillations-region-I}
\end{figure*}
In the reference cases discussed in the previous section, the interactions helped to induce transitions from stationary to power-oscillating operation regimes, but broke the relation between the two stationary solutions, while the power-oscillating states only set in with a vanishing amplitude and quickly became erratic as this amplitude increased. 

Balanced nonlinear particle interactions are repulsive ($g_{n,A}=g>0$) on one sublattice and attractive $g_{n,B}=-g<0$) on the other sublattice, and preserve the dynamical $\mathcal{PC}$ symmetry defined in Eq.~\eqref{eq:dynsym}. This combines the key features of the two reference cases. As we will show, we then encounter very different switching mechanisms, which lead to the emergence of a robust, symmetry-protected dynamical solution that sets in with a finite power-oscillation amplitude. We encounter two different scenarios for the transition, corresponding to the two regions I and II in Fig.~\ref{zero_phase_space}. Region I displays hysteresis and multistability between stationary and power-oscillating solutions, where the latter sets in with a finite oscillation period. Region II displays a clear division between stationary and power-oscillating solutions, where the latter now sets in with a initially diverging period. Because of these different dynamical and hysteretic signatures, the regions I and II can be interpreted as two different phases of the system. We therefore contrast our findings in the two regions, where we again study the solutions of the system as a function of increasing interaction strength $g$ for fixed values of the pump strength $\Gamma$ and decay rate $\gamma$.

\subsubsection{Region $I$}\label{sec:results2I}
As shown in Fig.~\ref{oscillations-region-I} for representative values $\gamma=0.015$, $\Gamma=0.5$ in region I, the two symmetry-related stationary solutions remain stable for finite balanced interaction strengths $g<g_{\mathrm{stat}}=0.00074$, where they are still related by the dynamical $\mathcal{PC}$ symmetry \eqref{eq:dynsym}. The interactions visibly distort the intensity distributions so that the two solutions are biased toward one of the two edges,
[see Fig. \ref{oscillations-region-I}(c,f)], and along with this have predominant population on one of the two sublattices [Fig. \ref{oscillations-region-I}(d,g)]. As conjugated solutions, these spatial distributions remain related by spatial reflection about the center of the system, while their frequency spectra remain related by reflection in the frequency domain [Fig. \ref{oscillations-region-I}(e,h)].

Both stationary solutions become simultaneously unstable above a critical interactions strength $g_{\mathrm{stat}}$. Already before this threshold is reached, however, the system also can sustain a time-dependent oscillating solution, which is stabilized  above a threshold $g_{\mathrm{osc}}=0.00065<g_{\mathrm{stat}}$, and hence coexists with the stationary solutions for interaction strengths
$g_{\mathrm{osc}}<g<g_{\mathrm{stat}}$. Remarkably, this oscillating solution is globally invariant under the transformation \eqref{eq:dynsym}, which translates it by half a period,
and therefore is protected by $\mathcal{PC}$-symmetry. The nonlinear interactions therefore induce a transition into a self-symmetric dynamical state, even though the noninteracting system does not allow this to happen.

As illustrated in Fig. \ref{oscillations-region-I}(i-k), the power oscillations are pronounced (up to $30\%$-$40\%$ of the total intensity) and are very robust, in that they survive for a wide range of values of $g$. Both the period and the oscillation amplitude are maximal at threshold and decrease for increasing values of $g$. This behavior can be observed by tracking an oscillating solution and gradually decreasing the strength of the nonlinearities. Moreover, the oscillations are characterized by a comb of frequencies where the dominant peaks decrease in number and increase in spacing when $g$ is increased. Resolved by position, the frequency spectrum of the A sublattice is symmetric to the spectrum of the B sublattice.

Physically these power-oscillations solutions originate from the competition of the positive energy shift on the sublattice with repulsive interaction and the negative energy shift on the sublattice with attractive interactions, which has to be taken into account when forming a state that makes optimal use of the gain on both sublattices.  Stronger interactions increase the energy gap between the two sublattices and consequently induce faster oscillations. Due to the specific structure of the nonlinear terms this effect can be achieved not only by increasing $g$, but also by increasing the pump intensity $\Gamma$, which is favourable for experiments.

A detailed investigation of the Bogoliubov spectra of the stationary solutions at threshold shows that the instability is caused by an excitation that moves along the imaginary axis from negative to positive imaginary values. This feature, together with the presence of a bistable region where two stable fixed points coexist with a limit cycle that sets in with a finite oscillation amplitude is characteristic of a saddle-node bifurcation of cycles \cite{Strogatz2015}.

\begin{figure}
\centering
\includegraphics[scale=0.285]{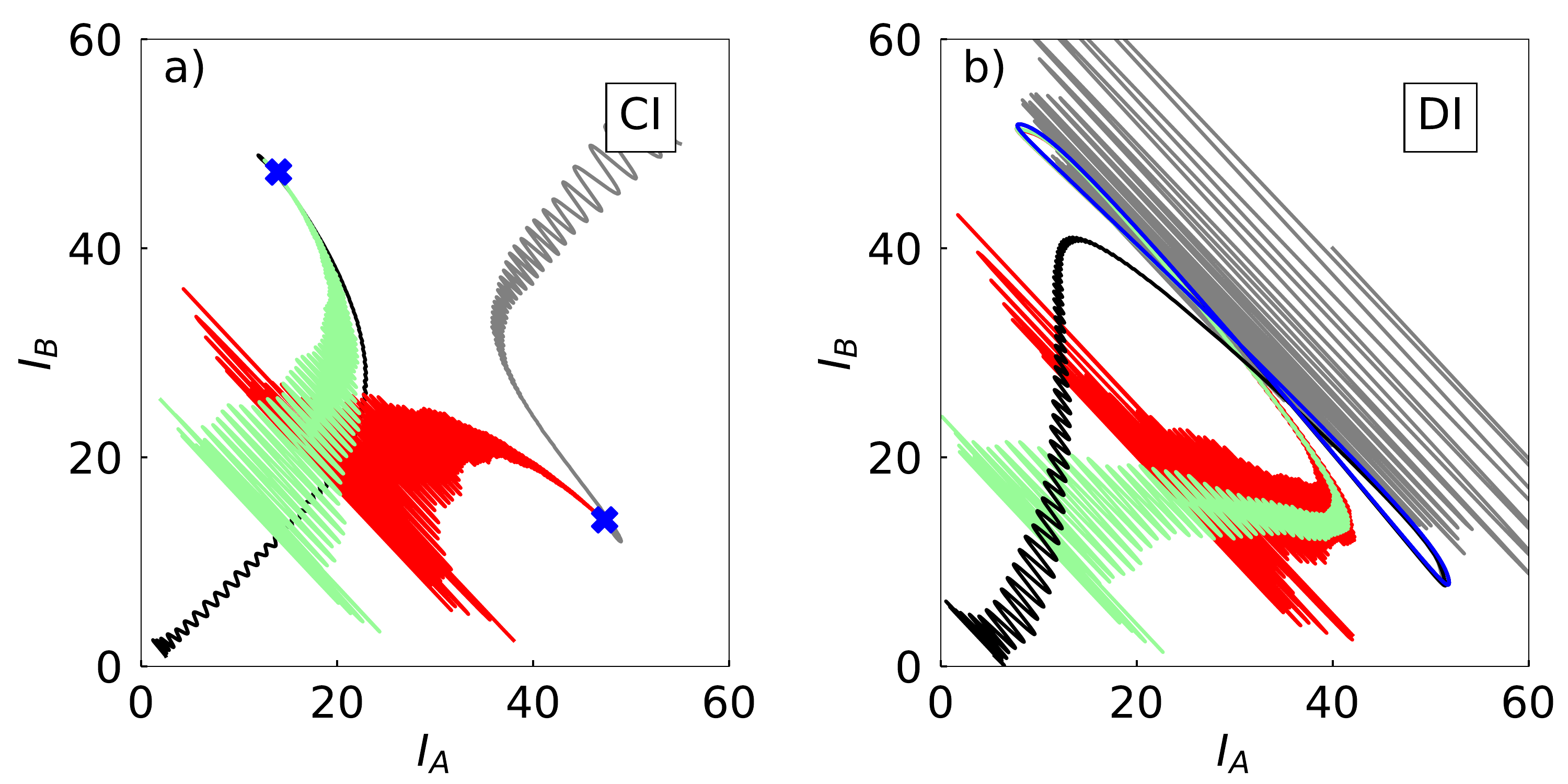}
\includegraphics[scale=0.285]{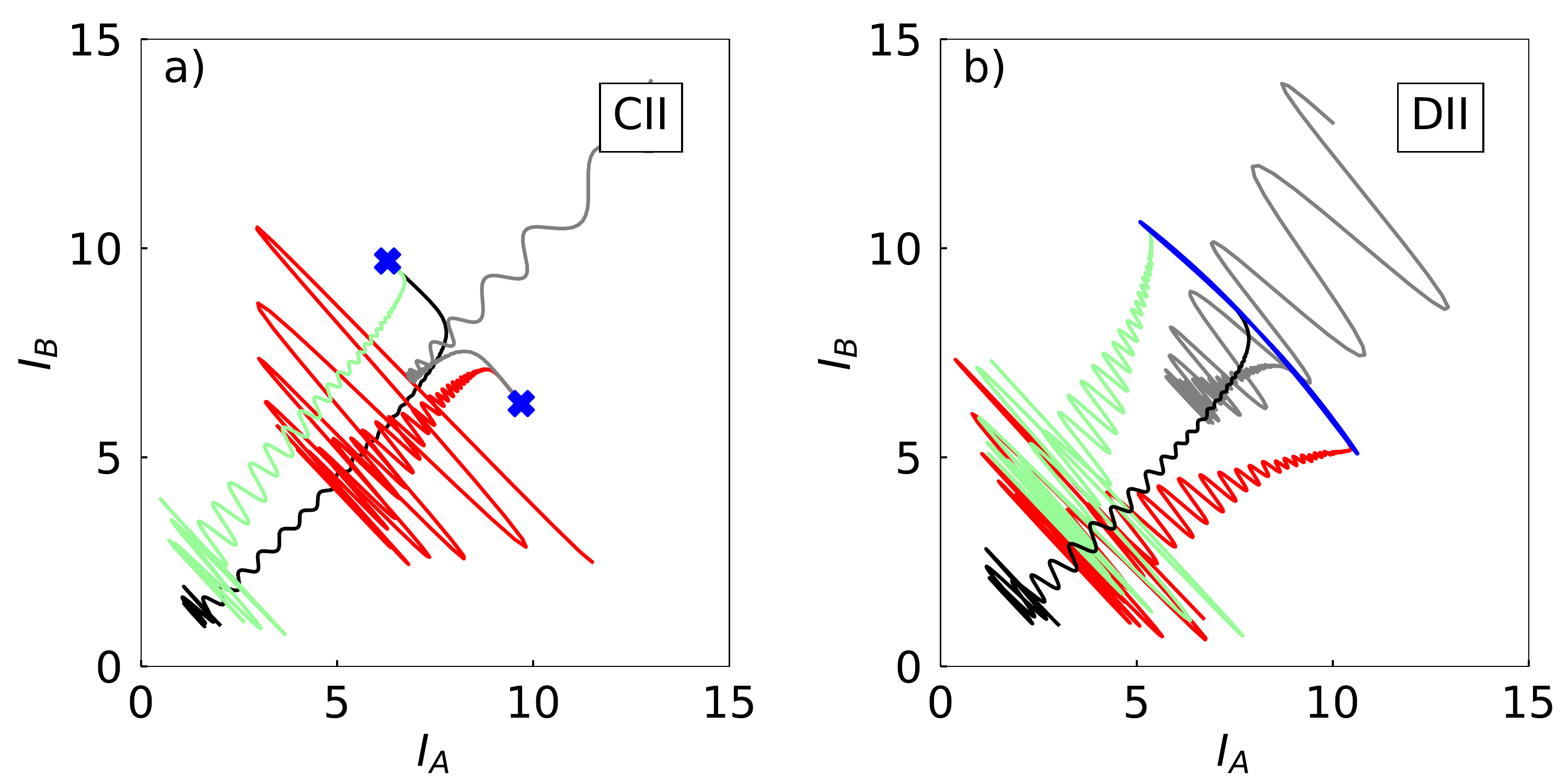}
\caption{{\bf Fixed points and limit cycles.} The time evolution of the system is plotted in the plane of the intensities $I_A(t)$ and $I_B(t)$ on the two sublattices for different initial conditions ($I_A(0),I_B(0)$). Panels (a) and (b) concern region I of Fig.~\ref{zero_phase_space} and correspond to points C and D in the bistable interval of Fig. \ref{oscillations-region-I}. In panel (a) the four different initial conditions $(2.5,1.0)$, $(2.0,25.5)$, $(38.0,2,5)$, $(55.0,50.0)$ (corresponding to the black, green, red, gray line respectively) lead to one of two fixed points  (blue crosses), which correspond to the stationary solutions,
while in panel (b) the four different initial conditions $(5.5,1.0)$, $(2.0,22.0)$, $(42.0,3.0)$, $(40.0,40.0)$,   lead to a limit cycle (in blue), which corresponds to the power-oscillating solution. Panel (c) and (d) concern region II and correspond to the two points C and D on either side of the power-oscillation threshold in Fig. \ref{oscillations-region-II}. In panel (c) the four different initial conditions $(2.0,1.0)$, $(0.5,4.0)$, $(11.5,2,5)$, $(13.0,14.0)$ lead to either one of the two fixed points (blue crosses), while in panel (d) all four different initial conditions $(3.0,1.0)$, $(1.3,7.3)$, $(6.7,1.15)$, $(10.0,13,0)$ lead to the limit cycle.}
\label{orbits}
\end{figure}

In Figure \ref{orbits} the time evolution of the system is plotted in the plane of the total intensities $I_A(t)$ and $I_B(t)$ on the two sublattices, where different lines correspond to different initial conditions. Panel (a) shows the case of four different initial conditions leading to one of the two fixed points (the two blue crosses), while panel (b) shows initial conditions leading to the limit cycle. These solutions  correspond to the points $C$ or $D$ in Fig.~\ref{oscillations-region-I}, respectively. 
For $g=0$ the two fixed points would correspond to the symmetric and anti-symmetric states and have identical intensities. Therefore, for $g=0$ the two blue crosses would lie on the line $I_A=I_B$. As the value of $g$ is increased the two fixed points move apart until they became unstable.

\subsubsection{Region $II$}
\begin{figure*}
\centering
\includegraphics[scale=0.33]{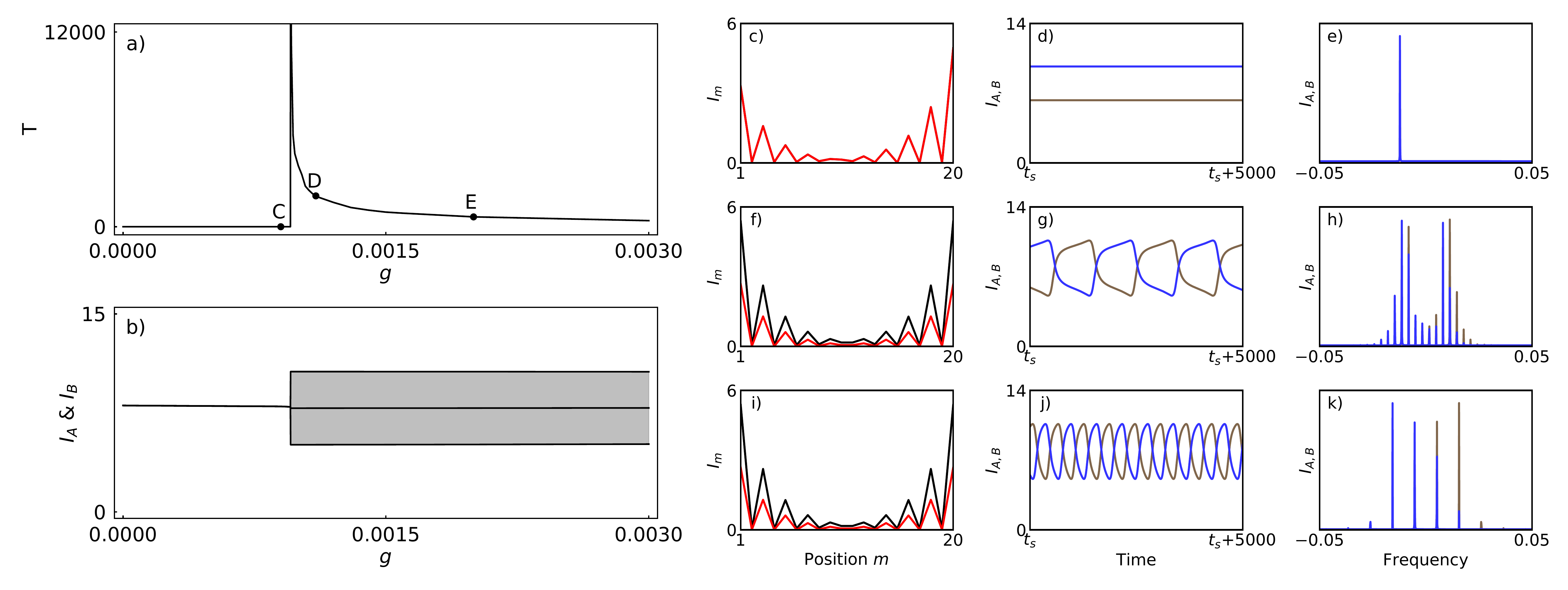}
\caption{{\bf Nonlinear pumped system with balanced interactions \boldmath$g_{n,A}=-g_{n,B}$
in region II.} Same as Fig.~\ref{oscillations-region-I}, but for parameters $\gamma=0.015$, and $\Gamma=0.5$ in region II. No bistable behaviour is observed and the period of the power oscillations diverges at threshold. Panels (c-e) show one of the two stationary solutions at point C ($g=0.0009$, the other solution is obtained from the dynamical symmetry \eqref{eq:dynsym}), while panels (f-h) and (i-k) show the behaviour of the power-oscillating solution at point D close to threshold ($g=0.0011$) and point E far above threshold ($g=0.0020$), respectively.}
\label{oscillations-region-II}
\end{figure*}

As already observed in Fig.~\ref{zero_phase_space}, in region II the stability excitation spectrum has undergone a transition so that the excitation that destabilized the stationary states in region I are shielded by two additional modes on the imaginary axis. The consequences for the dynamics of the system are illustrated in Fig.~\ref{oscillations-region-II}, where we again selected representative parameters for $\Gamma$ and $\gamma$. In contrast to the behaviour in region I, the system no longer exhibits a bistable interval where stationary states coexist with power-oscillating states. Instead, the emergence of the power-oscillating solution coincides with the instability threshold of the stationary states, which for the given parameters occurs at $g_{th}\approx 0.0009554$. Remarkably, at threshold the oscillation period now tends to infinity [see Fig. \ref{oscillations-region-II}(a)]. On the other hand, the two stationary solutions show the same general characteristics as in region I, so that one solution predominantly populates sublattice A  while the other populates sublattice B while the frequency spectra are symmetric to each other (Fig. \ref{oscillations-region-II}(c-e) shows one of these cases). Moreover, as before, the power-oscillations solutions are characterized by a frequency comb where the dominant peaks decrease in number with increasing $g$ [see Fig. \ref{oscillations-region-II}(h,k)]. Finally, increasing $g$ induces stronger nonlinear energy shifts between the two sublattices, which translates into faster oscillations [panels (g) and (j)] and wider spacing in the peaks of the frequency spectrum [panels (h) and (k)].

As before it is interesting to study the behaviour of the system at the threshold between the stationary and power oscillating solutions. The Bogoliubov spectrum of the two steady states confirms that the instability is still driven by an  excitation with zero real part moving up along the imaginary axis, but as already indicated above this excitation originates from the exceptional point that separates regions I and II. On the other side of the transition we observe that the oscillating period diverges to infinity as $(g-g_{th})^{-1/2}$. These two features, together with the finite oscillation amplitude at threshold, allow us to identify this transition as an infinite-period bifurcation \cite{Strogatz2015}. 

As for the case in region I we tracked the time evolution of the system in the plane of the intensity on sublattice $A$ and on sublattice $B$. In Fig.~\ref{orbits}(c), $g<g_{th}$ and 
all orbits converge to one of the two steady-state fixed points, while in Fig. \ref{orbits}(d) $g>g_{th}$ and all orbits converge to the oscillating limit cycle.
These solutions  correspond to the points $C$ or $D$ in Fig.~\ref{oscillations-region-II}, respectively. 

Finally we note that the transition between regions I and II is smooth. For the two cases considered above it is possible to move from one to the other by, e.g., increasing the homogeneous decay rate $\gamma$. In doing so the bistable region reduces in width and the  period of the power oscillations at threshold increases. At the boundary between the two regions the width of the bistable region goes to zero and the oscillation period to infinity. During this process the threshold value $g_{th}$ at which the power oscillations set in increases, since increasing the decay rate decreases the intensity in the system. The only effect of further increasing $\gamma$ after the region boundary is to further reduce the intensity and increase the threshold value $g_{th}$. Note that, alternatively, this transition can be performed  by increasing $\Gamma$, thereby moving left-right in the phase space of Fig.~\ref{zero_phase_space}, which may constitute a more suitable protocol in experimental realizations of the proposed system.

\section{Conclusions\label{sec:conclusions}}
In conclusion, we have investigated an SSH model in the presence of two types of nonlinearities: a laser-type nonlinearity describing a saturable pumping mechanism in the system, and a nonlinearity describing interactions that can be either attractive or repulsive. We identified a case of balanced interactions in which the system displays a novel combination of parity and charge-conjugation symmetries $\mathcal{PC}$, which extends dynamically to the nonlinear situation.

To explore the consequences of this symmetry we focused on the behavior of the edge states and studied the stability of stationary solutions as well as the emergence of power-oscillating solutions. We contrasted a scenario where uniform repulsive interactions on all lattice sites break the  $\mathcal{PC}$-symmetry  with a scenario  where the two sublattices displayed  balanced attractive and repulsive interactions so that the  $\mathcal{PC}$-symmetry is preserved.

The symmetry-broken case exhibits power-oscillating solutions that emerge in conventional Hopf bifurcations, hence set in with a finite period but vanishing oscillation amplitude, and furthermore develop chaotic behaviour when the interactions are increased.

In contrast,  the $\mathcal{PC}$-symmetric system supports power-oscillating solutions that are  globally invariant under the $\mathcal{PC}$-symmetry and hence turn out to be robust for large variations of the system parameters. These symmetry-protected power-oscillations
emerge via two distinct mechanisms, where in both cases the initial oscillation amplitude is finite, while the initial period can either be finite or divergent.

The latter feature suggests their implementation as a mediating state for an adiabatic switch between the two stationary edge-state solutions, which can be achieved by slowly increasing and decreasing the intensity of the pump, thus opening up new ways to design optical devices \cite{Ballarini2013,PhysRevB.92.174528}. The symmetry-protected power-oscillating states furthermore display a well-structured, symmetric frequency-comb spectrum that can be tuned by varying the pump or interaction strength. They thereby also naturally provide highly desirable features for the realization of coherent sources of terahertz radiation \cite{PhysRevLett.114.193901}.

\section*{Acknowledgments}
The work was supported by EPSRC Grant EP/N031776/1. The authors would like to thank S. Malzard, M. Szyniszewski , J. Arkinstall, G. Buonaiuto, and D. M. Whittaker for useful discussions and comments.

\appendix
\section{Bogoliubov excitation spectra\label{app:a}}
In this appendix we briefly review the definition and interpretation of the Bogoliubov excitation spectrum for a stationary state. For a compact presentation, we write
the coupled-mode equations \eqref{wave_equations} as
\begin{eqnarray}
i\frac{d\Psi}{dt}=[H_0+V(|\Psi|^2)]\Psi, \quad\Psi=\left(\begin{array}{c}A_1\\B_1\\A_2\\ \vdots \\ B_N\end{array}\right),
\label{eq:cmodeh}
\end{eqnarray}
where $\Psi$ is the vector grouping the wave amplitudes on the different sites. The Hamiltonian $H$ is an $N\times N$ matrix that includes the coupling terms $\tau$ and $\tau'$, while $V(|\Psi|^2)$ is a diagonal $N\times N$ matrix that includes the nonlinear interaction and saturation terms and the linear decay, as specified in Eq.~\eqref{eq:v}. For a stationary solution $\Psi(t)=\exp(-i\Omega t)\psi$ fulfilling $\Omega\psi=[H_0+V(|\psi|^2)]\psi$, the Bogoliubov analysis allows determining its stability against particle-like and hole-like perturbations $u$ and $v$ \cite{PhysRevB.82.224512,Malzard2018a},
\begin{equation}
\Psi(t)=[\psi+u \text{exp}(-i\omega t)+v^*\text{exp}(i\omega t)]\text{exp}(-i\Omega t).
\nonumber
\end{equation}
Inserting this expanded solution in the system of equations \eqref{eq:cmodeh} and linearising in $u$ and $v$ leads to the Bogoliubov equation
\begin{equation}
\omega\varphi=(\mathcal{H}[\psi]-\Sigma_z\Omega)\varphi,\quad\left(\begin{array}{c}u\\v \end{array}\right),
\label{eigenvalues}
\end{equation}
where $\Sigma_z$ is a Pauli matrix in the space of $u$ and $v$, and $\mathcal{H}[\psi]$ is the $2N\times2N$-dimensional Bogoliubov Hamiltonian
\begin{equation}
\mathcal{H}[\psi]=
\begin{bmatrix}
H_0+V+V'|\psi|^{2} & V'\psi^2\\
-[V'\psi^2]^* & -[H_0+V+V'|\psi|^2]^*\\
\end{bmatrix},
\nonumber
\end{equation}
where $V\equiv V(|\psi|^2)$ and
$V'\equiv\partial V(|\psi|^2)/\partial |\psi|^2$
are evaluated with the stationary solution.

In order to assess the stability of a steady state solution it is sufficient to evaluate the complex eigenvalues $\omega_n=\omega_n'+i\omega_n''$ of the Bogoliubov Hamiltonian, and study their imaginary part. We note that the Bogoliubov Hamiltonian always possesses one eigenvalue $\omega_0=0$, associated with $u_0=-v_0=\psi$. This describes the freedom to choose the global phase of the stationary state, and therefore constitutes a $U(1)$  Goldstone mode. All perturbations with negative imaginary parts $\omega_n''<0$ correspond to exponentially decaying perturbations, while positive imaginary parts corresponds to destabilizing perturbations that drive the system out of its steady state. For a stable stationary state, all perturbations besides the $U(1)$  Goldstone mode decay, and therefore the complete Bogoliubov spectrum lies in the lower half of the complex plane. A stationary state becomes unstable when eigenvalues cross the real axis. This constitutes a Hopf bifurcation if the instability occurs for eigenvalues with a finite real part $\omega_n'$, leading to an onset of oscillations with period $T=1/\omega_n'$.

\section{Comparison of symmetries\label{app:b}}
In the linear limit, the $\mathcal{PC}$-symmetry encountered in this work falls naturally into the Bernard-LeClair classification of nonhermitian systems \cite{BernardLeClair,PhysRevB.98.115135}. 
In this language, the system is characterized by transformation properties of the instantaneous Hamiltonian $H$ generating the time evolution (explicitly denoted as $H=H_0+V$ in App. \ref{app:a}), which here fulfills the condition $\mathcal{PC}H\mathcal{PC}=-H$ whilst the symmetry operator $\mathcal{PC}$ is antilinear and obeys $(\mathcal{PC})^2=-1$.

In general, this symmetry can be realized in systems with two subparts A and B (e.g. constituting two sublattices), which are coupled by an instantaneous Hamiltonian of the form
\begin{equation}
H=\left(\begin{array}{cc}V & T^*\\ T &  -V^* \end{array}\right). 
\end{equation}
We can naturally interpret  parity $\mathcal{P}=\sigma_x$ as the operation interchanging both subsystems (written as a Pauli matrix), and the charge conjugation  $\mathcal{C}=K\sigma_z$ as a combination of complex conjugation $K$ (for time reversal) and the standard chiral symmetry $\sigma_z$ of systems with two subparts.
In combination, this imposes the property $\sigma_y H\sigma_y=-H^*$.

As a consequence of this property the spectrum of the Hamiltonian consists of eigenvalues on the imaginary axis and pairs of eigenvalues that lie symmetrically to it. As we have seen, the complex couplings $T$ and $V$ may nonlinearly depend on the intensities along the system, upon which the spectral symmetry becomes the dynamical symmetry
\eqref{eq:dynsym}.

These features are similar, but distinct, to  various antilinear symmetries encountered in previous studies of nonhermitian photonic systems. First and foremost amongst these is the widely studied $\mathcal{PT}$ symmetry $\mathcal{PT}H\mathcal{PT}=H$ \cite{Bender,El-Ganainy:07}, 
where the parity operation is again typically realized by $\mathcal{P}=\sigma_x$ such that $(\mathcal{PT})^2=1$. Sending $H\to iH$ one can pass to 
anti-$\mathcal{PT}$-symmetric systems for which $\mathcal{PT}H\mathcal{PT}=-H$ \cite{PhysRevA.88.053810}.
Finally, by utilizing $\mathcal{C}=K\sigma_z$ with a matrix of possibly finite signature $\mathrm{tr}\,\sigma_z\neq 0$ one can realize a nonhermitian charge-conjugation symmetry $\mathcal{C}H\mathcal{C}=-H$ \cite{PhysRevLett.115.200402}, which admits topologically protected zero modes \cite{Schomerus:13,Poli2015} and also extends to the nonlinear setting \cite{Malzard2018a}.

\bibliography{biblio}

\begin{thebibliography}{47}%
\makeatletter
\providecommand \@ifxundefined [1]{%
 \@ifx{#1\undefined}
}%
\providecommand \@ifnum [1]{%
 \ifnum #1\expandafter \@firstoftwo
 \else \expandafter \@secondoftwo
 \fi
}%
\providecommand \@ifx [1]{%
 \ifx #1\expandafter \@firstoftwo
 \else \expandafter \@secondoftwo
 \fi
}%
\providecommand \natexlab [1]{#1}%
\providecommand \enquote  [1]{``#1''}%
\providecommand \bibnamefont  [1]{#1}%
\providecommand \bibfnamefont [1]{#1}%
\providecommand \citenamefont [1]{#1}%
\providecommand \href@noop [0]{\@secondoftwo}%
\providecommand \href [0]{\begingroup \@sanitize@url \@href}%
\providecommand \@href[1]{\@@startlink{#1}\@@href}%
\providecommand \@@href[1]{\endgroup#1\@@endlink}%
\providecommand \@sanitize@url [0]{\catcode `\\12\catcode `\$12\catcode
  `\&12\catcode `\#12\catcode `\^12\catcode `\_12\catcode `\%12\relax}%
\providecommand \@@startlink[1]{}%
\providecommand \@@endlink[0]{}%
\providecommand \url  [0]{\begingroup\@sanitize@url \@url }%
\providecommand \@url [1]{\endgroup\@href {#1}{\urlprefix }}%
\providecommand \urlprefix  [0]{URL }%
\providecommand \Eprint [0]{\href }%
\providecommand \doibase [0]{http://dx.doi.org/}%
\providecommand \selectlanguage [0]{\@gobble}%
\providecommand \bibinfo  [0]{\@secondoftwo}%
\providecommand \bibfield  [0]{\@secondoftwo}%
\providecommand \translation [1]{[#1]}%
\providecommand \BibitemOpen [0]{}%
\providecommand \bibitemStop [0]{}%
\providecommand \bibitemNoStop [0]{.\EOS\space}%
\providecommand \EOS [0]{\spacefactor3000\relax}%
\providecommand \BibitemShut  [1]{\csname bibitem#1\endcsname}%
\let\auto@bib@innerbib\@empty
\bibitem [{\citenamefont {{Ozawa}}\ \emph {et~al.}()\citenamefont {{Ozawa}},
  \citenamefont {{Price}}, \citenamefont {{Amo}}, \citenamefont {{Goldman}},
  \citenamefont {{Hafezi}}, \citenamefont {{Lu}}, \citenamefont {{Rechtsman}},
  \citenamefont {{Schuster}}, \citenamefont {{Simon}}, \citenamefont
  {{Zilberberg}},\ and\ \citenamefont {{Carusotto}}}]{2018arXiv180204173O}%
  \BibitemOpen
  \bibfield  {author} {\bibinfo {author} {\bibfnamefont {T.}~\bibnamefont
  {{Ozawa}}}, \bibinfo {author} {\bibfnamefont {H.~M.}\ \bibnamefont
  {{Price}}}, \bibinfo {author} {\bibfnamefont {A.}~\bibnamefont {{Amo}}},
  \bibinfo {author} {\bibfnamefont {N.}~\bibnamefont {{Goldman}}}, \bibinfo
  {author} {\bibfnamefont {M.}~\bibnamefont {{Hafezi}}}, \bibinfo {author}
  {\bibfnamefont {L.}~\bibnamefont {{Lu}}}, \bibinfo {author} {\bibfnamefont
  {M.}~\bibnamefont {{Rechtsman}}}, \bibinfo {author} {\bibfnamefont
  {D.}~\bibnamefont {{Schuster}}}, \bibinfo {author} {\bibfnamefont
  {J.}~\bibnamefont {{Simon}}}, \bibinfo {author} {\bibfnamefont
  {O.}~\bibnamefont {{Zilberberg}}}, \ and\ \bibinfo {author} {\bibfnamefont
  {I.}~\bibnamefont {{Carusotto}}},\ }\href@noop {} {\bibfield  {journal}
  {\bibinfo  {journal} {ArXiv}\ }}\Eprint {http://arxiv.org/abs/1802.04173}
  {1802.04173} \BibitemShut {NoStop}%
\bibitem [{\citenamefont {Lu}\ \emph {et~al.}(2014)\citenamefont {Lu},
  \citenamefont {Joannopoulos},\ and\ \citenamefont {Soljacic}}]{Lu2014}%
  \BibitemOpen
  \bibfield  {author} {\bibinfo {author} {\bibfnamefont {L.}~\bibnamefont
  {Lu}}, \bibinfo {author} {\bibfnamefont {J.~D.}\ \bibnamefont
  {Joannopoulos}}, \ and\ \bibinfo {author} {\bibfnamefont {M.}~\bibnamefont
  {Soljacic}},\ }\href {http://dx.doi.org/10.1038/nphoton.2014.248} {\bibfield
  {journal} {\bibinfo  {journal} {Nature Photonics}\ }\textbf {\bibinfo
  {volume} {8}},\ \bibinfo {pages} {821} (\bibinfo {year} {2014})}\BibitemShut
  {NoStop}%
\bibitem [{\citenamefont {Su}\ \emph {et~al.}(1979)\citenamefont {Su},
  \citenamefont {Schrieffer},\ and\ \citenamefont
  {Heeger}}]{PhysRevLett.42.1698}%
  \BibitemOpen
  \bibfield  {author} {\bibinfo {author} {\bibfnamefont {W.~P.}\ \bibnamefont
  {Su}}, \bibinfo {author} {\bibfnamefont {J.~R.}\ \bibnamefont {Schrieffer}},
  \ and\ \bibinfo {author} {\bibfnamefont {A.~J.}\ \bibnamefont {Heeger}},\
  }\href {\doibase 10.1103/PhysRevLett.42.1698} {\bibfield  {journal} {\bibinfo
   {journal} {Phys. Rev. Lett.}\ }\textbf {\bibinfo {volume} {42}},\ \bibinfo
  {pages} {1698} (\bibinfo {year} {1979})}\BibitemShut {NoStop}%
\bibitem [{\citenamefont {Malkova}\ \emph {et~al.}(2009)\citenamefont
  {Malkova}, \citenamefont {Hromada}, \citenamefont {Wang}, \citenamefont
  {Bryant},\ and\ \citenamefont {Chen}}]{Malkova:09}%
  \BibitemOpen
  \bibfield  {author} {\bibinfo {author} {\bibfnamefont {N.}~\bibnamefont
  {Malkova}}, \bibinfo {author} {\bibfnamefont {I.}~\bibnamefont {Hromada}},
  \bibinfo {author} {\bibfnamefont {X.}~\bibnamefont {Wang}}, \bibinfo {author}
  {\bibfnamefont {G.}~\bibnamefont {Bryant}}, \ and\ \bibinfo {author}
  {\bibfnamefont {Z.}~\bibnamefont {Chen}},\ }\href {\doibase
  10.1364/OL.34.001633} {\bibfield  {journal} {\bibinfo  {journal} {Opt.
  Lett.}\ }\textbf {\bibinfo {volume} {34}},\ \bibinfo {pages} {1633} (\bibinfo
  {year} {2009})}\BibitemShut {NoStop}%
\bibitem [{\citenamefont {Atala}\ \emph {et~al.}(2013)\citenamefont {Atala},
  \citenamefont {Aidelsburger}, \citenamefont {Barreiro}, \citenamefont
  {Abanin}, \citenamefont {Kitagawa}, \citenamefont {Demler},\ and\
  \citenamefont {Bloch}}]{Atala2013}%
  \BibitemOpen
  \bibfield  {author} {\bibinfo {author} {\bibfnamefont {M.}~\bibnamefont
  {Atala}}, \bibinfo {author} {\bibfnamefont {M.}~\bibnamefont {Aidelsburger}},
  \bibinfo {author} {\bibfnamefont {J.~T.}\ \bibnamefont {Barreiro}}, \bibinfo
  {author} {\bibfnamefont {D.}~\bibnamefont {Abanin}}, \bibinfo {author}
  {\bibfnamefont {T.}~\bibnamefont {Kitagawa}}, \bibinfo {author}
  {\bibfnamefont {E.}~\bibnamefont {Demler}}, \ and\ \bibinfo {author}
  {\bibfnamefont {I.}~\bibnamefont {Bloch}},\ }\href
  {http://dx.doi.org/10.1038/nphys2790} {\bibfield  {journal} {\bibinfo
  {journal} {Nature Physics}\ }\textbf {\bibinfo {volume} {9}},\ \bibinfo
  {pages} {795} (\bibinfo {year} {2013})}\BibitemShut {NoStop}%
\bibitem [{\citenamefont {Poli}\ \emph {et~al.}(2015)\citenamefont {Poli},
  \citenamefont {Bellec}, \citenamefont {Kuhl}, \citenamefont {Mortessagne},\
  and\ \citenamefont {Schomerus}}]{Poli2015}%
  \BibitemOpen
  \bibfield  {author} {\bibinfo {author} {\bibfnamefont {C.}~\bibnamefont
  {Poli}}, \bibinfo {author} {\bibfnamefont {M.}~\bibnamefont {Bellec}},
  \bibinfo {author} {\bibfnamefont {U.}~\bibnamefont {Kuhl}}, \bibinfo {author}
  {\bibfnamefont {F.}~\bibnamefont {Mortessagne}}, \ and\ \bibinfo {author}
  {\bibfnamefont {H.}~\bibnamefont {Schomerus}},\ }\href
  {http://dx.doi.org/10.1038/ncomms7710} {\bibfield  {journal} {\bibinfo
  {journal} {Nature Communications}\ }\textbf {\bibinfo {volume} {6}},\
  \bibinfo {pages} {6710} (\bibinfo {year} {2015})}\BibitemShut {NoStop}%
\bibitem [{\citenamefont {Zhao}\ \emph {et~al.}(2018)\citenamefont {Zhao},
  \citenamefont {Miao}, \citenamefont {Teimourpour}, \citenamefont {Malzard},
  \citenamefont {El-Ganainy}, \citenamefont {Schomerus},\ and\ \citenamefont
  {Feng}}]{Zhao2018}%
  \BibitemOpen
  \bibfield  {author} {\bibinfo {author} {\bibfnamefont {H.}~\bibnamefont
  {Zhao}}, \bibinfo {author} {\bibfnamefont {P.}~\bibnamefont {Miao}}, \bibinfo
  {author} {\bibfnamefont {M.~H.}\ \bibnamefont {Teimourpour}}, \bibinfo
  {author} {\bibfnamefont {S.}~\bibnamefont {Malzard}}, \bibinfo {author}
  {\bibfnamefont {R.}~\bibnamefont {El-Ganainy}}, \bibinfo {author}
  {\bibfnamefont {H.}~\bibnamefont {Schomerus}}, \ and\ \bibinfo {author}
  {\bibfnamefont {L.}~\bibnamefont {Feng}},\ }\href {\doibase
  10.1038/s41467-018-03434-2} {\bibfield  {journal} {\bibinfo  {journal}
  {Nature Communications}\ }\textbf {\bibinfo {volume} {9}},\ \bibinfo {pages}
  {981} (\bibinfo {year} {2018})}\BibitemShut {NoStop}%
\bibitem [{\citenamefont {Parto}\ \emph {et~al.}(2018)\citenamefont {Parto},
  \citenamefont {Wittek}, \citenamefont {Hodaei}, \citenamefont {Harari},
  \citenamefont {Bandres}, \citenamefont {Ren}, \citenamefont {Rechtsman},
  \citenamefont {Segev}, \citenamefont {Christodoulides},\ and\ \citenamefont
  {Khajavikhan}}]{PhysRevLett.120.113901}%
  \BibitemOpen
  \bibfield  {author} {\bibinfo {author} {\bibfnamefont {M.}~\bibnamefont
  {Parto}}, \bibinfo {author} {\bibfnamefont {S.}~\bibnamefont {Wittek}},
  \bibinfo {author} {\bibfnamefont {H.}~\bibnamefont {Hodaei}}, \bibinfo
  {author} {\bibfnamefont {G.}~\bibnamefont {Harari}}, \bibinfo {author}
  {\bibfnamefont {M.~A.}\ \bibnamefont {Bandres}}, \bibinfo {author}
  {\bibfnamefont {J.}~\bibnamefont {Ren}}, \bibinfo {author} {\bibfnamefont
  {M.~C.}\ \bibnamefont {Rechtsman}}, \bibinfo {author} {\bibfnamefont
  {M.}~\bibnamefont {Segev}}, \bibinfo {author} {\bibfnamefont {D.~N.}\
  \bibnamefont {Christodoulides}}, \ and\ \bibinfo {author} {\bibfnamefont
  {M.}~\bibnamefont {Khajavikhan}},\ }\href {\doibase
  10.1103/PhysRevLett.120.113901} {\bibfield  {journal} {\bibinfo  {journal}
  {Phys. Rev. Lett.}\ }\textbf {\bibinfo {volume} {120}},\ \bibinfo {pages}
  {113901} (\bibinfo {year} {2018})}\BibitemShut {NoStop}%
\bibitem [{\citenamefont {{Yao}}\ \emph {et~al.}()\citenamefont {{Yao}},
  \citenamefont {{Li}}, \citenamefont {{Zheng}}, \citenamefont {{An}},
  \citenamefont {{Ding}}, \citenamefont {{Lee}}, \citenamefont {{Zhang}},\ and\
  \citenamefont {{Guo}}}]{2018arXiv180401587Y}%
  \BibitemOpen
  \bibfield  {author} {\bibinfo {author} {\bibfnamefont {R.}~\bibnamefont
  {{Yao}}}, \bibinfo {author} {\bibfnamefont {H.}~\bibnamefont {{Li}}},
  \bibinfo {author} {\bibfnamefont {B.}~\bibnamefont {{Zheng}}}, \bibinfo
  {author} {\bibfnamefont {S.}~\bibnamefont {{An}}}, \bibinfo {author}
  {\bibfnamefont {J.}~\bibnamefont {{Ding}}}, \bibinfo {author} {\bibfnamefont
  {C.-S.}\ \bibnamefont {{Lee}}}, \bibinfo {author} {\bibfnamefont
  {H.}~\bibnamefont {{Zhang}}}, \ and\ \bibinfo {author} {\bibfnamefont
  {W.}~\bibnamefont {{Guo}}},\ }\href@noop {} {\bibfield  {journal} {\bibinfo
  {journal} {ArXiv}\ }}\Eprint {http://arxiv.org/abs/1804.01587} {1804.01587}
  \BibitemShut {NoStop}%
\bibitem [{\citenamefont {{Ota}}\ \emph {et~al.}()\citenamefont {{Ota}},
  \citenamefont {{Katsumi}}, \citenamefont {{Watanabe}}, \citenamefont
  {{Iwamoto}},\ and\ \citenamefont {{Arakawa}}}]{2018arXiv180609826O}%
  \BibitemOpen
  \bibfield  {author} {\bibinfo {author} {\bibfnamefont {Y.}~\bibnamefont
  {{Ota}}}, \bibinfo {author} {\bibfnamefont {R.}~\bibnamefont {{Katsumi}}},
  \bibinfo {author} {\bibfnamefont {K.}~\bibnamefont {{Watanabe}}}, \bibinfo
  {author} {\bibfnamefont {S.}~\bibnamefont {{Iwamoto}}}, \ and\ \bibinfo
  {author} {\bibfnamefont {Y.}~\bibnamefont {{Arakawa}}},\ }\href@noop {}
  {\bibfield  {journal} {\bibinfo  {journal} {ArXiv}\ }}\Eprint
  {http://arxiv.org/abs/1806.09826} {1806.09826} \BibitemShut {NoStop}%
\bibitem [{\citenamefont {Zeuner}\ \emph {et~al.}(2015)\citenamefont {Zeuner},
  \citenamefont {Rechtsman}, \citenamefont {Plotnik}, \citenamefont {Lumer},
  \citenamefont {Nolte}, \citenamefont {Rudner}, \citenamefont {Segev},\ and\
  \citenamefont {Szameit}}]{PhysRevLett.115.040402}%
  \BibitemOpen
  \bibfield  {author} {\bibinfo {author} {\bibfnamefont {J.~M.}\ \bibnamefont
  {Zeuner}}, \bibinfo {author} {\bibfnamefont {M.~C.}\ \bibnamefont
  {Rechtsman}}, \bibinfo {author} {\bibfnamefont {Y.}~\bibnamefont {Plotnik}},
  \bibinfo {author} {\bibfnamefont {Y.}~\bibnamefont {Lumer}}, \bibinfo
  {author} {\bibfnamefont {S.}~\bibnamefont {Nolte}}, \bibinfo {author}
  {\bibfnamefont {M.~S.}\ \bibnamefont {Rudner}}, \bibinfo {author}
  {\bibfnamefont {M.}~\bibnamefont {Segev}}, \ and\ \bibinfo {author}
  {\bibfnamefont {A.}~\bibnamefont {Szameit}},\ }\href {\doibase
  10.1103/PhysRevLett.115.040402} {\bibfield  {journal} {\bibinfo  {journal}
  {Phys. Rev. Lett.}\ }\textbf {\bibinfo {volume} {115}},\ \bibinfo {pages}
  {040402} (\bibinfo {year} {2015})}\BibitemShut {NoStop}%
\bibitem [{\citenamefont {Bleckmann}\ \emph {et~al.}(2017)\citenamefont
  {Bleckmann}, \citenamefont {Cherpakova}, \citenamefont {Linden},\ and\
  \citenamefont {Alberti}}]{PhysRevB.96.045417}%
  \BibitemOpen
  \bibfield  {author} {\bibinfo {author} {\bibfnamefont {F.}~\bibnamefont
  {Bleckmann}}, \bibinfo {author} {\bibfnamefont {Z.}~\bibnamefont
  {Cherpakova}}, \bibinfo {author} {\bibfnamefont {S.}~\bibnamefont {Linden}},
  \ and\ \bibinfo {author} {\bibfnamefont {A.}~\bibnamefont {Alberti}},\ }\href
  {\doibase 10.1103/PhysRevB.96.045417} {\bibfield  {journal} {\bibinfo
  {journal} {Phys. Rev. B}\ }\textbf {\bibinfo {volume} {96}},\ \bibinfo
  {pages} {045417} (\bibinfo {year} {2017})}\BibitemShut {NoStop}%
\bibitem [{\citenamefont {St-Jean}\ \emph {et~al.}(2017)\citenamefont
  {St-Jean}, \citenamefont {Goblot}, \citenamefont {Galopin}, \citenamefont
  {Lema\^{i}tre}, \citenamefont {Ozawa}, \citenamefont {Le~Gratiet},
  \citenamefont {Sagnes}, \citenamefont {Bloch},\ and\ \citenamefont
  {Amo}}]{St-Jean2017}%
  \BibitemOpen
  \bibfield  {author} {\bibinfo {author} {\bibfnamefont {P.}~\bibnamefont
  {St-Jean}}, \bibinfo {author} {\bibfnamefont {V.}~\bibnamefont {Goblot}},
  \bibinfo {author} {\bibfnamefont {E.}~\bibnamefont {Galopin}}, \bibinfo
  {author} {\bibfnamefont {A.}~\bibnamefont {Lema\^{i}tre}}, \bibinfo {author}
  {\bibfnamefont {T.}~\bibnamefont {Ozawa}}, \bibinfo {author} {\bibfnamefont
  {L.}~\bibnamefont {Le~Gratiet}}, \bibinfo {author} {\bibfnamefont
  {I.}~\bibnamefont {Sagnes}}, \bibinfo {author} {\bibfnamefont
  {J.}~\bibnamefont {Bloch}}, \ and\ \bibinfo {author} {\bibfnamefont
  {A.}~\bibnamefont {Amo}},\ }\href {\doibase 10.1038/s41566-017-0006-2}
  {\bibfield  {journal} {\bibinfo  {journal} {Nature Photonics}\ }\textbf
  {\bibinfo {volume} {11}},\ \bibinfo {pages} {651} (\bibinfo {year}
  {2017})}\BibitemShut {NoStop}%
\bibitem [{\citenamefont {Malzard}\ and\ \citenamefont
  {Schomerus}(2018)}]{Malzard2018}%
  \BibitemOpen
  \bibfield  {author} {\bibinfo {author} {\bibfnamefont {S.}~\bibnamefont
  {Malzard}}\ and\ \bibinfo {author} {\bibfnamefont {H.}~\bibnamefont
  {Schomerus}},\ }\href {http://stacks.iop.org/1367-2630/20/i=6/a=063044}
  {\bibfield  {journal} {\bibinfo  {journal} {New Journal of Physics}\ }\textbf
  {\bibinfo {volume} {20}},\ \bibinfo {pages} {063044} (\bibinfo {year}
  {2018})}\BibitemShut {NoStop}%
\bibitem [{\citenamefont {Malzard}\ \emph {et~al.}(2018)\citenamefont
  {Malzard}, \citenamefont {Cancellieri},\ and\ \citenamefont
  {Schomerus}}]{Malzard2018a}%
  \BibitemOpen
  \bibfield  {author} {\bibinfo {author} {\bibfnamefont {S.}~\bibnamefont
  {Malzard}}, \bibinfo {author} {\bibfnamefont {E.}~\bibnamefont
  {Cancellieri}}, \ and\ \bibinfo {author} {\bibfnamefont {H.}~\bibnamefont
  {Schomerus}},\ }\href {\doibase 10.1364/OE.26.022506} {\bibfield  {journal}
  {\bibinfo  {journal} {Opt. Express}\ }\textbf {\bibinfo {volume} {26}},\
  \bibinfo {pages} {22506} (\bibinfo {year} {2018})}\BibitemShut {NoStop}%
\bibitem [{\citenamefont {Webb}\ and\ \citenamefont
  {Hooker}(2010)}]{laser_textbook}%
  \BibitemOpen
  \bibfield  {author} {\bibinfo {author} {\bibfnamefont {C.}~\bibnamefont
  {Webb}}\ and\ \bibinfo {author} {\bibfnamefont {S.}~\bibnamefont {Hooker}},\
  }\href@noop {} {\emph {\bibinfo {title} {Laser Physics}}}\ (\bibinfo
  {publisher} {Oxford University Press},\ \bibinfo {year} {2010})\BibitemShut
  {NoStop}%
\bibitem [{\citenamefont {Zhou}\ and\ \citenamefont {Chong}(2016)}]{Zhou:16}%
  \BibitemOpen
  \bibfield  {author} {\bibinfo {author} {\bibfnamefont {X.}~\bibnamefont
  {Zhou}}\ and\ \bibinfo {author} {\bibfnamefont {Y.~D.}\ \bibnamefont
  {Chong}},\ }\href {\doibase 10.1364/OE.24.006916} {\bibfield  {journal}
  {\bibinfo  {journal} {Opt. Express}\ }\textbf {\bibinfo {volume} {24}},\
  \bibinfo {pages} {6916} (\bibinfo {year} {2016})}\BibitemShut {NoStop}%
\bibitem [{\citenamefont {Hadad}\ \emph {et~al.}(2016)\citenamefont {Hadad},
  \citenamefont {Khanikaev},\ and\ \citenamefont {Al\`u}}]{PhysRevB.93.155112}%
  \BibitemOpen
  \bibfield  {author} {\bibinfo {author} {\bibfnamefont {Y.}~\bibnamefont
  {Hadad}}, \bibinfo {author} {\bibfnamefont {A.~B.}\ \bibnamefont
  {Khanikaev}}, \ and\ \bibinfo {author} {\bibfnamefont {A.}~\bibnamefont
  {Al\`u}},\ }\href {\doibase 10.1103/PhysRevB.93.155112} {\bibfield  {journal}
  {\bibinfo  {journal} {Phys. Rev. B}\ }\textbf {\bibinfo {volume} {93}},\
  \bibinfo {pages} {155112} (\bibinfo {year} {2016})}\BibitemShut {NoStop}%
\bibitem [{\citenamefont {Zhou}\ \emph {et~al.}(2017)\citenamefont {Zhou},
  \citenamefont {Wang}, \citenamefont {Leykam},\ and\ \citenamefont
  {Chong}}]{1367-2630-19-9-095002}%
  \BibitemOpen
  \bibfield  {author} {\bibinfo {author} {\bibfnamefont {X.}~\bibnamefont
  {Zhou}}, \bibinfo {author} {\bibfnamefont {Y.}~\bibnamefont {Wang}}, \bibinfo
  {author} {\bibfnamefont {D.}~\bibnamefont {Leykam}}, \ and\ \bibinfo {author}
  {\bibfnamefont {Y.~D.}\ \bibnamefont {Chong}},\ }\href
  {http://stacks.iop.org/1367-2630/19/i=9/a=095002} {\bibfield  {journal}
  {\bibinfo  {journal} {New Journal of Physics}\ }\textbf {\bibinfo {volume}
  {19}},\ \bibinfo {pages} {095002} (\bibinfo {year} {2017})}\BibitemShut
  {NoStop}%
\bibitem [{\citenamefont {Li}\ \emph {et~al.}(2017)\citenamefont {Li},
  \citenamefont {Hu}, \citenamefont {Yang},\ and\ \citenamefont
  {Gong}}]{doi:10.1063/1.4976013}%
  \BibitemOpen
  \bibfield  {author} {\bibinfo {author} {\bibfnamefont {C.}~\bibnamefont
  {Li}}, \bibinfo {author} {\bibfnamefont {X.}~\bibnamefont {Hu}}, \bibinfo
  {author} {\bibfnamefont {H.}~\bibnamefont {Yang}}, \ and\ \bibinfo {author}
  {\bibfnamefont {Q.}~\bibnamefont {Gong}},\ }\href {\doibase
  10.1063/1.4976013} {\bibfield  {journal} {\bibinfo  {journal} {AIP Advances}\
  }\textbf {\bibinfo {volume} {7}},\ \bibinfo {pages} {025203} (\bibinfo {year}
  {2017})}\BibitemShut {NoStop}%
\bibitem [{\citenamefont {Lumer}\ \emph {et~al.}(2013)\citenamefont {Lumer},
  \citenamefont {Plotnik}, \citenamefont {Rechtsman},\ and\ \citenamefont
  {Segev}}]{PhysRevLett.111.243905}%
  \BibitemOpen
  \bibfield  {author} {\bibinfo {author} {\bibfnamefont {Y.}~\bibnamefont
  {Lumer}}, \bibinfo {author} {\bibfnamefont {Y.}~\bibnamefont {Plotnik}},
  \bibinfo {author} {\bibfnamefont {M.~C.}\ \bibnamefont {Rechtsman}}, \ and\
  \bibinfo {author} {\bibfnamefont {M.}~\bibnamefont {Segev}},\ }\href
  {\doibase 10.1103/PhysRevLett.111.243905} {\bibfield  {journal} {\bibinfo
  {journal} {Phys. Rev. Lett.}\ }\textbf {\bibinfo {volume} {111}},\ \bibinfo
  {pages} {243905} (\bibinfo {year} {2013})}\BibitemShut {NoStop}%
\bibitem [{\citenamefont {Carusotto}\ and\ \citenamefont
  {Ciuti}(2013)}]{RevModPhys.85.299}%
  \BibitemOpen
  \bibfield  {author} {\bibinfo {author} {\bibfnamefont {I.}~\bibnamefont
  {Carusotto}}\ and\ \bibinfo {author} {\bibfnamefont {C.}~\bibnamefont
  {Ciuti}},\ }\href {\doibase 10.1103/RevModPhys.85.299} {\bibfield  {journal}
  {\bibinfo  {journal} {Rev. Mod. Phys.}\ }\textbf {\bibinfo {volume} {85}},\
  \bibinfo {pages} {299} (\bibinfo {year} {2013})}\BibitemShut {NoStop}%
\bibitem [{\citenamefont {Barnett}(2013)}]{PhysRevA.88.063631}%
  \BibitemOpen
  \bibfield  {author} {\bibinfo {author} {\bibfnamefont {R.}~\bibnamefont
  {Barnett}},\ }\href {\doibase 10.1103/PhysRevA.88.063631} {\bibfield
  {journal} {\bibinfo  {journal} {Phys. Rev. A}\ }\textbf {\bibinfo {volume}
  {88}},\ \bibinfo {pages} {063631} (\bibinfo {year} {2013})}\BibitemShut
  {NoStop}%
\bibitem [{\citenamefont {Di~Liberto}\ \emph {et~al.}(2016)\citenamefont
  {Di~Liberto}, \citenamefont {Recati}, \citenamefont {Carusotto},\ and\
  \citenamefont {Menotti}}]{PhysRevA.94.062704}%
  \BibitemOpen
  \bibfield  {author} {\bibinfo {author} {\bibfnamefont {M.}~\bibnamefont
  {Di~Liberto}}, \bibinfo {author} {\bibfnamefont {A.}~\bibnamefont {Recati}},
  \bibinfo {author} {\bibfnamefont {I.}~\bibnamefont {Carusotto}}, \ and\
  \bibinfo {author} {\bibfnamefont {C.}~\bibnamefont {Menotti}},\ }\href
  {\doibase 10.1103/PhysRevA.94.062704} {\bibfield  {journal} {\bibinfo
  {journal} {Phys. Rev. A}\ }\textbf {\bibinfo {volume} {94}},\ \bibinfo
  {pages} {062704} (\bibinfo {year} {2016})}\BibitemShut {NoStop}%
\bibitem [{\citenamefont {Dobrykh}\ \emph {et~al.}()\citenamefont {Dobrykh},
  \citenamefont {Yulin}, \citenamefont {Slobozhanyuk}, \citenamefont
  {Poddubny},\ and\ \citenamefont {Kivshar}}]{1805.03760}%
  \BibitemOpen
  \bibfield  {author} {\bibinfo {author} {\bibfnamefont {D.~A.}\ \bibnamefont
  {Dobrykh}}, \bibinfo {author} {\bibfnamefont {A.~V.}\ \bibnamefont {Yulin}},
  \bibinfo {author} {\bibfnamefont {A.~P.}\ \bibnamefont {Slobozhanyuk}},
  \bibinfo {author} {\bibfnamefont {A.~N.}\ \bibnamefont {Poddubny}}, \ and\
  \bibinfo {author} {\bibfnamefont {Y.~S.}\ \bibnamefont {Kivshar}},\
  }\href@noop {} {\enquote {\bibinfo {title} {Observation and control of
  nonlinear electromagnetic topological edge states},}\ }\Eprint
  {http://arxiv.org/abs/arXiv:1805.03760} {arXiv:1805.03760} \BibitemShut
  {NoStop}%
\bibitem [{\citenamefont {Gulevich}\ \emph {et~al.}(2017)\citenamefont
  {Gulevich}, \citenamefont {Yudin}, \citenamefont {Skryabin}, \citenamefont
  {Iorsh},\ and\ \citenamefont {Shelykh}}]{Gulevich2017}%
  \BibitemOpen
  \bibfield  {author} {\bibinfo {author} {\bibfnamefont {D.~R.}\ \bibnamefont
  {Gulevich}}, \bibinfo {author} {\bibfnamefont {D.}~\bibnamefont {Yudin}},
  \bibinfo {author} {\bibfnamefont {D.~V.}\ \bibnamefont {Skryabin}}, \bibinfo
  {author} {\bibfnamefont {I.~V.}\ \bibnamefont {Iorsh}}, \ and\ \bibinfo
  {author} {\bibfnamefont {I.~A.}\ \bibnamefont {Shelykh}},\ }\href {\doibase
  10.1038/s41598-017-01646-y} {\bibfield  {journal} {\bibinfo  {journal}
  {Scientific Reports}\ }\textbf {\bibinfo {volume} {7}},\ \bibinfo {pages}
  {1780} (\bibinfo {year} {2017})}\BibitemShut {NoStop}%
\bibitem [{\citenamefont {Ablowitz}\ \emph {et~al.}(2014)\citenamefont
  {Ablowitz}, \citenamefont {Curtis},\ and\ \citenamefont
  {Ma}}]{PhysRevA.90.023813}%
  \BibitemOpen
  \bibfield  {author} {\bibinfo {author} {\bibfnamefont {M.~J.}\ \bibnamefont
  {Ablowitz}}, \bibinfo {author} {\bibfnamefont {C.~W.}\ \bibnamefont
  {Curtis}}, \ and\ \bibinfo {author} {\bibfnamefont {Y.-P.}\ \bibnamefont
  {Ma}},\ }\href {\doibase 10.1103/PhysRevA.90.023813} {\bibfield  {journal}
  {\bibinfo  {journal} {Phys. Rev. A}\ }\textbf {\bibinfo {volume} {90}},\
  \bibinfo {pages} {023813} (\bibinfo {year} {2014})}\BibitemShut {NoStop}%
\bibitem [{\citenamefont {Leykam}\ and\ \citenamefont
  {Chong}(2016)}]{PhysRevLett.117.143901}%
  \BibitemOpen
  \bibfield  {author} {\bibinfo {author} {\bibfnamefont {D.}~\bibnamefont
  {Leykam}}\ and\ \bibinfo {author} {\bibfnamefont {Y.~D.}\ \bibnamefont
  {Chong}},\ }\href {\doibase 10.1103/PhysRevLett.117.143901} {\bibfield
  {journal} {\bibinfo  {journal} {Phys. Rev. Lett.}\ }\textbf {\bibinfo
  {volume} {117}},\ \bibinfo {pages} {143901} (\bibinfo {year}
  {2016})}\BibitemShut {NoStop}%
\bibitem [{\citenamefont {Suntsov}\ \emph {et~al.}(2006)\citenamefont
  {Suntsov}, \citenamefont {Makris}, \citenamefont {Christodoulides},
  \citenamefont {Stegeman}, \citenamefont {Hach\'e}, \citenamefont
  {Morandotti}, \citenamefont {Yang}, \citenamefont {Salamo},\ and\
  \citenamefont {Sorel}}]{PhysRevLett.96.063901}%
  \BibitemOpen
  \bibfield  {author} {\bibinfo {author} {\bibfnamefont {S.}~\bibnamefont
  {Suntsov}}, \bibinfo {author} {\bibfnamefont {K.~G.}\ \bibnamefont {Makris}},
  \bibinfo {author} {\bibfnamefont {D.~N.}\ \bibnamefont {Christodoulides}},
  \bibinfo {author} {\bibfnamefont {G.~I.}\ \bibnamefont {Stegeman}}, \bibinfo
  {author} {\bibfnamefont {A.}~\bibnamefont {Hach\'e}}, \bibinfo {author}
  {\bibfnamefont {R.}~\bibnamefont {Morandotti}}, \bibinfo {author}
  {\bibfnamefont {H.}~\bibnamefont {Yang}}, \bibinfo {author} {\bibfnamefont
  {G.}~\bibnamefont {Salamo}}, \ and\ \bibinfo {author} {\bibfnamefont
  {M.}~\bibnamefont {Sorel}},\ }\href {\doibase 10.1103/PhysRevLett.96.063901}
  {\bibfield  {journal} {\bibinfo  {journal} {Phys. Rev. Lett.}\ }\textbf
  {\bibinfo {volume} {96}},\ \bibinfo {pages} {063901} (\bibinfo {year}
  {2006})}\BibitemShut {NoStop}%
\bibitem [{\citenamefont {Suntsov}\ \emph {et~al.}(2007)\citenamefont
  {Suntsov}, \citenamefont {Makris}, \citenamefont {Christodoulides},
  \citenamefont {Stegeman}, \citenamefont {Morandotti}, \citenamefont {Yang},
  \citenamefont {Salamo},\ and\ \citenamefont {Sorel}}]{Suntsov:07}%
  \BibitemOpen
  \bibfield  {author} {\bibinfo {author} {\bibfnamefont {S.}~\bibnamefont
  {Suntsov}}, \bibinfo {author} {\bibfnamefont {K.~G.}\ \bibnamefont {Makris}},
  \bibinfo {author} {\bibfnamefont {D.~N.}\ \bibnamefont {Christodoulides}},
  \bibinfo {author} {\bibfnamefont {G.~I.}\ \bibnamefont {Stegeman}}, \bibinfo
  {author} {\bibfnamefont {R.}~\bibnamefont {Morandotti}}, \bibinfo {author}
  {\bibfnamefont {H.}~\bibnamefont {Yang}}, \bibinfo {author} {\bibfnamefont
  {G.}~\bibnamefont {Salamo}}, \ and\ \bibinfo {author} {\bibfnamefont
  {M.}~\bibnamefont {Sorel}},\ }\href {\doibase 10.1364/OL.32.003098}
  {\bibfield  {journal} {\bibinfo  {journal} {Opt. Lett.}\ }\textbf {\bibinfo
  {volume} {32}},\ \bibinfo {pages} {3098} (\bibinfo {year}
  {2007})}\BibitemShut {NoStop}%
\bibitem [{\citenamefont {Lederer}\ \emph {et~al.}(2008)\citenamefont
  {Lederer}, \citenamefont {Stegeman}, \citenamefont {Christodoulides},
  \citenamefont {Assanto}, \citenamefont {Segev},\ and\ \citenamefont
  {Silberberg}}]{LEDERER20081}%
  \BibitemOpen
  \bibfield  {author} {\bibinfo {author} {\bibfnamefont {F.}~\bibnamefont
  {Lederer}}, \bibinfo {author} {\bibfnamefont {G.~I.}\ \bibnamefont
  {Stegeman}}, \bibinfo {author} {\bibfnamefont {D.~N.}\ \bibnamefont
  {Christodoulides}}, \bibinfo {author} {\bibfnamefont {G.}~\bibnamefont
  {Assanto}}, \bibinfo {author} {\bibfnamefont {M.}~\bibnamefont {Segev}}, \
  and\ \bibinfo {author} {\bibfnamefont {Y.}~\bibnamefont {Silberberg}},\
  }\href {\doibase https://doi.org/10.1016/j.physrep.2008.04.004} {\bibfield
  {journal} {\bibinfo  {journal} {Physics Reports}\ }\textbf {\bibinfo {volume}
  {463}},\ \bibinfo {pages} {1 } (\bibinfo {year} {2008})}\BibitemShut
  {NoStop}%
\bibitem [{\citenamefont {Bender}(2007)}]{Bender}%
  \BibitemOpen
  \bibfield  {author} {\bibinfo {author} {\bibfnamefont {C.~M.}\ \bibnamefont
  {Bender}},\ }\href {http://stacks.iop.org/0034-4885/70/i=6/a=R03} {\bibfield
  {journal} {\bibinfo  {journal} {Reports on Progress in Physics}\ }\textbf
  {\bibinfo {volume} {70}},\ \bibinfo {pages} {947} (\bibinfo {year}
  {2007})}\BibitemShut {NoStop}%
\bibitem [{\citenamefont {El-Ganainy}\ \emph {et~al.}(2007)\citenamefont
  {El-Ganainy}, \citenamefont {Makris}, \citenamefont {Christodoulides},\ and\
  \citenamefont {Musslimani}}]{El-Ganainy:07}%
  \BibitemOpen
  \bibfield  {author} {\bibinfo {author} {\bibfnamefont {R.}~\bibnamefont
  {El-Ganainy}}, \bibinfo {author} {\bibfnamefont {K.~G.}\ \bibnamefont
  {Makris}}, \bibinfo {author} {\bibfnamefont {D.~N.}\ \bibnamefont
  {Christodoulides}}, \ and\ \bibinfo {author} {\bibfnamefont {Z.~H.}\
  \bibnamefont {Musslimani}},\ }\href {\doibase 10.1364/OL.32.002632}
  {\bibfield  {journal} {\bibinfo  {journal} {Opt. Lett.}\ }\textbf {\bibinfo
  {volume} {32}},\ \bibinfo {pages} {2632} (\bibinfo {year}
  {2007})}\BibitemShut {NoStop}%
\bibitem [{\citenamefont {Moiseyev}(2011)}]{moiseyev_2011}%
  \BibitemOpen
  \bibfield  {author} {\bibinfo {author} {\bibfnamefont {N.}~\bibnamefont
  {Moiseyev}},\ }\href {\doibase 10.1017/CBO9780511976186} {\emph {\bibinfo
  {title} {Non-Hermitian Quantum Mechanics}}}\ (\bibinfo  {publisher}
  {Cambridge University Press},\ \bibinfo {year} {2011})\BibitemShut {NoStop}%
\bibitem [{\citenamefont {Malzard}\ \emph {et~al.}(2015)\citenamefont
  {Malzard}, \citenamefont {Poli},\ and\ \citenamefont
  {Schomerus}}]{PhysRevLett.115.200402}%
  \BibitemOpen
  \bibfield  {author} {\bibinfo {author} {\bibfnamefont {S.}~\bibnamefont
  {Malzard}}, \bibinfo {author} {\bibfnamefont {C.}~\bibnamefont {Poli}}, \
  and\ \bibinfo {author} {\bibfnamefont {H.}~\bibnamefont {Schomerus}},\ }\href
  {\doibase 10.1103/PhysRevLett.115.200402} {\bibfield  {journal} {\bibinfo
  {journal} {Phys. Rev. Lett.}\ }\textbf {\bibinfo {volume} {115}},\ \bibinfo
  {pages} {200402} (\bibinfo {year} {2015})}\BibitemShut {NoStop}%
\bibitem [{\citenamefont {Strogarz}(2015)}]{Strogatz2015}%
  \BibitemOpen
  \bibfield  {author} {\bibinfo {author} {\bibfnamefont {S.}~\bibnamefont
  {Strogarz}},\ }\href@noop {} {\emph {\bibinfo {title} {Nonlinear Dynamics and
  Chaos}}}\ (\bibinfo  {publisher} {Boca Raton: CRC Press},\ \bibinfo {year}
  {2015})\BibitemShut {NoStop}%
\bibitem [{\citenamefont {Giordmaine}\ and\ \citenamefont
  {Miller}(1965)}]{PhysRevLett.14.973}%
  \BibitemOpen
  \bibfield  {author} {\bibinfo {author} {\bibfnamefont {J.~A.}\ \bibnamefont
  {Giordmaine}}\ and\ \bibinfo {author} {\bibfnamefont {R.~C.}\ \bibnamefont
  {Miller}},\ }\href {\doibase 10.1103/PhysRevLett.14.973} {\bibfield
  {journal} {\bibinfo  {journal} {Phys. Rev. Lett.}\ }\textbf {\bibinfo
  {volume} {14}},\ \bibinfo {pages} {973} (\bibinfo {year} {1965})}\BibitemShut
  {NoStop}%
\bibitem [{\citenamefont {Whittaker}(2005)}]{PhysRevB.71.115301}%
  \BibitemOpen
  \bibfield  {author} {\bibinfo {author} {\bibfnamefont {D.~M.}\ \bibnamefont
  {Whittaker}},\ }\href {\doibase 10.1103/PhysRevB.71.115301} {\bibfield
  {journal} {\bibinfo  {journal} {Phys. Rev. B}\ }\textbf {\bibinfo {volume}
  {71}},\ \bibinfo {pages} {115301} (\bibinfo {year} {2005})}\BibitemShut
  {NoStop}%
\bibitem [{\citenamefont {Berceanu}\ \emph {et~al.}(2015)\citenamefont
  {Berceanu}, \citenamefont {Dominici}, \citenamefont {Carusotto},
  \citenamefont {Ballarini}, \citenamefont {Cancellieri}, \citenamefont
  {Gigli}, \citenamefont {Szyma\ifmmode~\acute{n}\else \'{n}\fi{}ska},
  \citenamefont {Sanvitto},\ and\ \citenamefont
  {Marchetti}}]{PhysRevB.92.035307}%
  \BibitemOpen
  \bibfield  {author} {\bibinfo {author} {\bibfnamefont {A.~C.}\ \bibnamefont
  {Berceanu}}, \bibinfo {author} {\bibfnamefont {L.}~\bibnamefont {Dominici}},
  \bibinfo {author} {\bibfnamefont {I.}~\bibnamefont {Carusotto}}, \bibinfo
  {author} {\bibfnamefont {D.}~\bibnamefont {Ballarini}}, \bibinfo {author}
  {\bibfnamefont {E.}~\bibnamefont {Cancellieri}}, \bibinfo {author}
  {\bibfnamefont {G.}~\bibnamefont {Gigli}}, \bibinfo {author} {\bibfnamefont
  {M.~H.}\ \bibnamefont {Szyma\ifmmode~\acute{n}\else \'{n}\fi{}ska}}, \bibinfo
  {author} {\bibfnamefont {D.}~\bibnamefont {Sanvitto}}, \ and\ \bibinfo
  {author} {\bibfnamefont {F.~M.}\ \bibnamefont {Marchetti}},\ }\href {\doibase
  10.1103/PhysRevB.92.035307} {\bibfield  {journal} {\bibinfo  {journal} {Phys.
  Rev. B}\ }\textbf {\bibinfo {volume} {92}},\ \bibinfo {pages} {035307}
  (\bibinfo {year} {2015})}\BibitemShut {NoStop}%
\bibitem [{\citenamefont {Rayanov}\ \emph {et~al.}(2015)\citenamefont
  {Rayanov}, \citenamefont {Altshuler}, \citenamefont {Rubo},\ and\
  \citenamefont {Flach}}]{PhysRevLett.114.193901}%
  \BibitemOpen
  \bibfield  {author} {\bibinfo {author} {\bibfnamefont {K.}~\bibnamefont
  {Rayanov}}, \bibinfo {author} {\bibfnamefont {B.~L.}\ \bibnamefont
  {Altshuler}}, \bibinfo {author} {\bibfnamefont {Y.~G.}\ \bibnamefont {Rubo}},
  \ and\ \bibinfo {author} {\bibfnamefont {S.}~\bibnamefont {Flach}},\ }\href
  {\doibase 10.1103/PhysRevLett.114.193901} {\bibfield  {journal} {\bibinfo
  {journal} {Phys. Rev. Lett.}\ }\textbf {\bibinfo {volume} {114}},\ \bibinfo
  {pages} {193901} (\bibinfo {year} {2015})}\BibitemShut {NoStop}%
\bibitem [{\citenamefont {Ballarini}\ \emph {et~al.}(2013)\citenamefont
  {Ballarini}, \citenamefont {De~Giorgi}, \citenamefont {Cancellieri},
  \citenamefont {Houdr{\'e}}, \citenamefont {Giacobino}, \citenamefont
  {Cingolani}, \citenamefont {Bramati}, \citenamefont {Gigli},\ and\
  \citenamefont {Sanvitto}}]{Ballarini2013}%
  \BibitemOpen
  \bibfield  {author} {\bibinfo {author} {\bibfnamefont {D.}~\bibnamefont
  {Ballarini}}, \bibinfo {author} {\bibfnamefont {M.}~\bibnamefont
  {De~Giorgi}}, \bibinfo {author} {\bibfnamefont {E.}~\bibnamefont
  {Cancellieri}}, \bibinfo {author} {\bibfnamefont {R.}~\bibnamefont
  {Houdr{\'e}}}, \bibinfo {author} {\bibfnamefont {E.}~\bibnamefont
  {Giacobino}}, \bibinfo {author} {\bibfnamefont {R.}~\bibnamefont
  {Cingolani}}, \bibinfo {author} {\bibfnamefont {A.}~\bibnamefont {Bramati}},
  \bibinfo {author} {\bibfnamefont {G.}~\bibnamefont {Gigli}}, \ and\ \bibinfo
  {author} {\bibfnamefont {D.}~\bibnamefont {Sanvitto}},\ }\href
  {http://dx.doi.org/10.1038/ncomms2734} {\bibfield  {journal} {\bibinfo
  {journal} {Nature Communications}\ }\textbf {\bibinfo {volume} {4}},\
  \bibinfo {pages} {1778} (\bibinfo {year} {2013})}\BibitemShut {NoStop}%
\bibitem [{\citenamefont {Cancellieri}\ \emph {et~al.}(2015)\citenamefont
  {Cancellieri}, \citenamefont {Chana}, \citenamefont {Sich}, \citenamefont
  {Krizhanovskii}, \citenamefont {Skolnick},\ and\ \citenamefont
  {Whittaker}}]{PhysRevB.92.174528}%
  \BibitemOpen
  \bibfield  {author} {\bibinfo {author} {\bibfnamefont {E.}~\bibnamefont
  {Cancellieri}}, \bibinfo {author} {\bibfnamefont {J.~K.}\ \bibnamefont
  {Chana}}, \bibinfo {author} {\bibfnamefont {M.}~\bibnamefont {Sich}},
  \bibinfo {author} {\bibfnamefont {D.~N.}\ \bibnamefont {Krizhanovskii}},
  \bibinfo {author} {\bibfnamefont {M.~S.}\ \bibnamefont {Skolnick}}, \ and\
  \bibinfo {author} {\bibfnamefont {D.~M.}\ \bibnamefont {Whittaker}},\ }\href
  {\doibase 10.1103/PhysRevB.92.174528} {\bibfield  {journal} {\bibinfo
  {journal} {Phys. Rev. B}\ }\textbf {\bibinfo {volume} {92}},\ \bibinfo
  {pages} {174528} (\bibinfo {year} {2015})}\BibitemShut {NoStop}%
\bibitem [{\citenamefont {Cancellieri}\ \emph {et~al.}(2010)\citenamefont
  {Cancellieri}, \citenamefont {Marchetti}, \citenamefont
  {Szyma\ifmmode~\acute{n}\else \'{n}\fi{}ska},\ and\ \citenamefont
  {Tejedor}}]{PhysRevB.82.224512}%
  \BibitemOpen
  \bibfield  {author} {\bibinfo {author} {\bibfnamefont {E.}~\bibnamefont
  {Cancellieri}}, \bibinfo {author} {\bibfnamefont {F.~M.}\ \bibnamefont
  {Marchetti}}, \bibinfo {author} {\bibfnamefont {M.~H.}\ \bibnamefont
  {Szyma\ifmmode~\acute{n}\else \'{n}\fi{}ska}}, \ and\ \bibinfo {author}
  {\bibfnamefont {C.}~\bibnamefont {Tejedor}},\ }\href {\doibase
  10.1103/PhysRevB.82.224512} {\bibfield  {journal} {\bibinfo  {journal} {Phys.
  Rev. B}\ }\textbf {\bibinfo {volume} {82}},\ \bibinfo {pages} {224512}
  (\bibinfo {year} {2010})}\BibitemShut {NoStop}%
\bibitem [{\citenamefont {{Bernard}}\ and\ \citenamefont
  {{LeClair}}()}]{BernardLeClair}%
  \BibitemOpen
  \bibfield  {author} {\bibinfo {author} {\bibfnamefont {D.}~\bibnamefont
  {{Bernard}}}\ and\ \bibinfo {author} {\bibfnamefont {A.}~\bibnamefont
  {{LeClair}}},\ }\href@noop {} {\bibinfo  {journal} {arXiv:0110649}\
  }\BibitemShut {NoStop}%
\bibitem [{\citenamefont {Lieu}(2018)}]{PhysRevB.98.115135}%
  \BibitemOpen
\bibfield  {journal} {  }\bibfield  {author} {\bibinfo {author} {\bibfnamefont
  {S.}~\bibnamefont {Lieu}},\ }\href {\doibase 10.1103/PhysRevB.98.115135}
  {\bibfield  {journal} {\bibinfo  {journal} {Phys. Rev. B}\ }\textbf {\bibinfo
  {volume} {98}},\ \bibinfo {pages} {115135} (\bibinfo {year}
  {2018})}\BibitemShut {NoStop}%
\bibitem [{\citenamefont {Ge}\ and\ \citenamefont
  {T\"ureci}(2013)}]{PhysRevA.88.053810}%
  \BibitemOpen
  \bibfield  {author} {\bibinfo {author} {\bibfnamefont {L.}~\bibnamefont
  {Ge}}\ and\ \bibinfo {author} {\bibfnamefont {H.~E.}\ \bibnamefont
  {T\"ureci}},\ }\href {\doibase 10.1103/PhysRevA.88.053810} {\bibfield
  {journal} {\bibinfo  {journal} {Phys. Rev. A}\ }\textbf {\bibinfo {volume}
  {88}},\ \bibinfo {pages} {053810} (\bibinfo {year} {2013})}\BibitemShut
  {NoStop}%
\bibitem [{\citenamefont {Schomerus}(2013)}]{Schomerus:13}%
  \BibitemOpen
  \bibfield  {author} {\bibinfo {author} {\bibfnamefont {H.}~\bibnamefont
  {Schomerus}},\ }\href {\doibase 10.1364/OL.38.001912} {\bibfield  {journal}
  {\bibinfo  {journal} {Opt. Lett.}\ }\textbf {\bibinfo {volume} {38}},\
  \bibinfo {pages} {1912} (\bibinfo {year} {2013})}\BibitemShut {NoStop}%
\end{thebibliography}%

\end{document}